%% file: neurips_2026.tex
\documentclass{article}

\PassOptionsToPackage{numbers, compress}{natbib}

\usepackage[eandd,preprint]{neurips_2026}

\usepackage[table]{xcolor}
\usepackage{listings}
\usepackage{pifont}
\usepackage{wrapfig}
\usepackage{multirow}
\usepackage{ulem}
\usepackage{pgfplots}
\usepackage{algorithm}
\usepackage{amssymb}
\usepackage{algpseudocode}

\pgfplotsset{compat=1.18} 

\newcommand{\cmark}{\ding{51}}
\newcommand{\xmark}{\ding{55}}
\definecolor{pykeyword}{RGB}{0,0,255}
\definecolor{pycomment}{RGB}{0,128,0}
\definecolor{pystring}{RGB}{163,21,21}
\definecolor{linegray}{RGB}{200,200,200}

\usepackage[utf8]{inputenc} 
\usepackage[T1]{fontenc}    
\usepackage{hyperref}       
\usepackage{url}            
\usepackage{booktabs}       
\usepackage{amsfonts}       
\usepackage{amsmath}        
\usepackage{nicefrac}       
\usepackage{microtype}      
\usepackage{xcolor}         
\usepackage{graphicx}       
\usepackage{subcaption}     
\usepackage{multirow}       
\usepackage{enumitem}       
\usepackage{amsmath}
\lstdefinestyle{base_style}{
    basicstyle=\ttfamily\footnotesize,
    breaklines=true,
    breakindent=0pt,             
    breakatwhitespace=false,
    xleftmargin=0pt,             
    xrightmargin=0pt,
    frame=lines,                 
    framerule=0.5pt,
    rulecolor=\color{linegray},
    keepspaces=true,
    showstringspaces=false,
    aboveskip=10pt,
    belowskip=10pt,
    columns=flexible
}

\newcommand{\TODO}[1][]{\textcolor{red}{[TODO\ifx&#1&\else: #1\fi]}}

\title{SWE-Cycle: Benchmarking Code Agents across the Complete Issue Resolution Cycle}

\author{
  \textbf{Hao Guan\textsuperscript{1}$^{*,\diamond}$},
  \textbf{Lingyue Fu\textsuperscript{1}$^{*}$},
  \textbf{Shao Zhang\textsuperscript{1}},
  \textbf{Yaoming Zhu\textsuperscript{2}},
  \textbf{Kangning Zhang\textsuperscript{1}},
  \textbf{Lin Qiu\textsuperscript{2}},
  \\
  \textbf{Xunliang Cai\textsuperscript{2}},
  \textbf{Xuezhi Cao\textsuperscript{2}},
  \textbf{Weiwen Liu\textsuperscript{1}},
  \textbf{Weinan Zhang\textsuperscript{1}},
  \textbf{Yong Yu\textsuperscript{1}}
\\
  \textsuperscript{1}Shanghai Jiao Tong University,
  \textsuperscript{2}Meituan,
  \\
  \small{
  $^*$ Equal contribution \quad
    $^{\diamond}$ Work done while interning at Meituan
  }
  \\
}

\begin{document}

\maketitle

\begin{abstract}
As autonomous code agents move toward end-to-end software development, evaluating their practical autonomy becomes critical. Current benchmarks hide friction by testing agents in pre-configured environments, and their static evaluation pipelines frequently fail when parsing fully autonomous trajectories. We address these limitations with SWE-Cycle, a benchmark of 489 rigorously filtered instances. SWE-Cycle evaluates agents across three isolated tasks, including environment reconstruction, code implementation, and verification test generation, as well as an end-to-end FullCycle task that integrates all three. The FullCycle task requires agents to work autonomously in a bare repository without human scaffolding.
To reliably assess these complex execution paths, we developed SWE-Judge. By combining static code review with dynamic testing, this execution-capable evaluation agent accurately verifies functional correctness and eliminates the systematic measurement errors of traditional static parsers.
We evaluate code agents powered by six state-of-the-art LLMs across these four tasks. The results reveal a sharp drop in solve rates when transitioning from isolated tasks to FullCycle execution, exposing critical bottlenecks in handling cross-phase dependencies and maintaining code quality. Together, SWE-Cycle and SWE-Judge provide a comprehensive framework for accurately measuring the end-to-end capabilities of autonomous software agents.

\end{abstract}

\section{Introduction}
\label{sec:intro}
The continuous enhancement of large language models (LLMs) in tool invocation and complex reasoning~\citep{anthropic2025claude46, glm5team2026glm5}, alongside the rapid evolution of agentic frameworks~\citep{opencode2026, anthropic2025claudecode, codex}, has driven significant progress in autonomous code agents. Modern code agents have transitioned from completing isolated algorithmic functions to maintaining entire engineering-level projects. By navigating large legacy codebases and synthesizing cross-file edits, they can now execute continuous development and integrate into real-world software pipelines~\citep{liu2025cicd}. As these code agents evolve into autonomous developers, their expanding capabilities require robust evaluation paradigms that accurately capture their practical autonomy.

Evaluation frameworks quantify this progress across distinct paradigms. Some benchmarks~\citep{jimenez2024swebench, chowdhury2024swebenchverified, deng2025swebenchpro, zan2025multiswebench} focus on issue resolution within existing codebases, requiring code agents to generate verified source code patches across diverse programming languages and complexities. Beyond patching, another paradigm assesses the full software development lifecycle through greenfield project creation, where agents execute system design, module implementation, and testing~\citep{golnari2026devbenchrealisticdeveloperinformedbenchmark, zeng2025e2edevbench}. Additionally, a third category targets specific development stages, evaluating code agents on configuring execution environments~\citep{eliseeva2025envbench} and generating automated test cases~\citep{wang2025testevalbenchmarkinglargelanguage}. Across these categories, execution-based metrics verify the functional correctness and runtime stability of the generated outputs. While recent evaluations cover a broader range of software engineering tasks, current benchmarks still fail to capture autonomous execution across the complete issue resolution lifecycle and rely on brittle, static evaluation pipelines.

Existing frameworks structurally fragment the development pipeline and bypass the complexity of legacy codebases~\citep{zeng2025e2edevbench, golnari2026devbenchrealisticdeveloperinformedbenchmark, ding2026nl2repobenchlonghorizonrepositorygeneration}, thereby failing to reflect an agent's true autonomy in real-world maintenance. Successfully resolving a real-world issue requires a unified progression: reconstructing the execution environment, implementing the fix, and generating verification tests. When these interrelated phases are artificially isolated~\citep{eliseeva2025envbench,wang2025testevalbenchmarkinglargelanguage}, evaluations shield code agents from cascading errors and the friction of full-process integration, such as dependency resolution and version control navigation~\citep{cemri2025multiagentfail}. Consequently, current benchmarks project an illusion of capability: high performance on pre-configured tasks severely overestimates true autonomy, masking the reality that agents frequently fail completely when forced to onboard codebases, implement fixes, and verify their work without intervention.

Moreover, static execution pipelines based on predefined unit tests and rigid parsers introduce systematic measurement errors and collapse in fully autonomous workflows. These deterministic systems are highly brittle. At the validation level, predefined static tests often suffer from flawed ground truths, such as misaligned checks or underspecified assertions~\citep{chowdhury2024swebenchverified, zhu2025establishing}. At the extraction level, strict parsers frequently misclassify functionally correct code due to trivial formatting deviations. Consequently, these rigid scripts frequently penalize valid alternative implementations and erroneously overlook deep logical flaws. More critically, this static paradigm is structurally incompatible with end-to-end scenarios. Because predefined scoring scripts cannot adapt to a code agent's dynamic and autonomous behaviors, traditional execution evaluation is fundamentally inadequate for assessing the complete issue resolution cycle.

To address these limitations, we present \textbf{SWE-Cycle}, a benchmark evaluating code agents across the complete issue resolution lifecycle. We construct this benchmark from SWE-bench Verified, Pro, and Multilingual, rigorously filtering them to retain 489 high-quality instances. Mapping directly to real-world software engineering, each instance integrates three essential tasks: environment reconstruction, code implementation, and verification test generation.
SWE-Cycle supports two evaluation settings. In the Isolated Task setting, each phase is evaluated independently with the remaining stages fully provided, enabling controlled comparisons with existing benchmarks. In the FullCycle setting, the agent receives only a bare repository and an issue description, requiring it to autonomously complete all three stages without human scaffolding.

To overcome the structural failures of rigid execution protocols, we further propose \textbf{SWE-Judge}, a hybrid evaluation agent integrating static code review with dynamic execution. Moving beyond the binary pass/fail metrics of predefined scripts, SWE-Judge employs task-specific validation protocols to accommodate diverse valid implementations, uncover structural defects, and capture fine-grained partial correctness. We systematically evaluate code agents powered by six state-of-the-art LLMs, reporting comprehensive results across both the isolated tasks and the FullCycle task. Our empirical analysis reveals that even within Isolated Tasks, traditional deterministic script evaluation produces severe misjudgments and false signals. In contrast, SWE-Judge provides a reliable assessment protocol across all four SWE-Cycle tasks and is strongly validated against human annotations. Ultimately, SWE-Cycle and SWE-Judge establish a unified benchmark and evaluation framework for assessing code agents across the complete issue resolution lifecycle.

In summary, our key contributions are as follows:
\begin{itemize}[leftmargin=10pt]
\item We introduce \textbf{SWE-Cycle}, the first benchmark evaluating agents across the complete issue resolution lifecycle. Curated via a rigorous filtering pipeline to retain 489 high-quality instances, it supports both independent Isolated Task evaluation and a FullCycle setting that requires agents to operate autonomously from a bare repository.

\item We propose \textbf{SWE-Judge}, the only evaluation paradigm capable of assessing the complete issue resolution tasks. Integrating static code review with dynamic execution, SWE-Judge overcomes the limitations of rigid scripts to accommodate diverse valid implementations, uncover structural defects, and capture fine-grained partial correctness.

\item We systematically evaluate six LLMs across both the isolated tasks and the FullCycle task. This extensive evaluation establishes a comprehensive capability profile, explicitly measuring model proficiency in environment reconstruction, code implementation, and verification test generation to provide a holistic view of their true autonomous potential.

\end{itemize}

\section{Related Work}
\label{sec:related}

\textbf{Code Agent Benchmarks.}
SWE-bench~\citep{jimenez2024swebench} has established the standard for evaluating code agents on real-world GitHub issues. This benchmark has spurred the rapid development of specialized agent architectures~\citep{yang2024sweagent, wang2024openhands}. Subsequent variants refine the evaluation along multiple dimensions: human-verified instance quality~\citep{chowdhury2024swebenchverified}, longer-horizon tasks~\citep{deng2025swebenchpro}, multilingual coverage~\citep{zan2025multiswebench}, continuous updates~\citep{gu2025swebenchlive}, heterogeneous comprehensive tasks~\citep{xu2025swecompassunifiedevaluationagentic}, and large-scale training data synthesis~\citep{yang2025swesmith}. Despite these diverse advances, all SWE-bench variants inherit the same structural limitation: they supply pre-built Docker environments and evaluate agents against fixed gold test suites. This design entirely excludes environment reconstruction and test generation from the evaluation scope.

Alternative benchmarks address orthogonal aspects of software engineering, evaluating agents on greenfield project creation~\citep{zeng2025e2edevbench}, feature development~\citep{li2025feabench, zhou2026featurebench}, or automated code review~\citep{zhang2026ccrab}. However, greenfield development differs fundamentally from maintaining legacy systems; real-world engineering predominantly involves navigating complex existing codebases rather than starting from empty directories. Meanwhile, EnvBench~\citep{eliseeva2025envbench} isolates environment reconstruction as a standalone capability, completely disconnecting it from downstream code implementation and verification test generation. Consequently, no existing benchmark captures the complete issue resolution lifecycle within a unified progression.

\textbf{Automated Evaluation Approaches.} Existing code agent benchmarks predominantly rely on unit test execution, which suffers from well-documented limitations: flawed gold tests, binary pass/fail scoring that discards partial correctness, and inapplicability when agents must generate their own verification code~\citep{chowdhury2024swebenchverified}. LLM-as-a-judge offers an alternative approach that assesses output quality on continuous scales without requiring gold references~\citep{zheng2024judging, gu2024surveyLLMjudge}. However, systematic studies reveal significant reliability concerns, including position bias that causes agreement fluctuations of up to 14\%~\citep{raina2025positionbias} and the fundamental limitations of execution-free judges when verifying runtime behavior~\citep{liu2024codejudgebench}. To address these shortcomings, Agent-as-a-Judge represents an emerging paradigm that augments LLM judges with agentic capabilities, including planning, tool use, and iterative verification~\citep{you2026surveyagentjudge}. Recent works demonstrate that such agentic evaluators, especially when fine-tuned, substantially outperform pure LLM-as-a-judge methods in both human alignment and cost efficiency~\citep{zhuge2025agentasjudge, wang2025prdbench}. SWE-Cycle adopts this paradigm through SWE-Judge, an execution-capable evaluation agent that combines static code review with dynamic test execution to overcome the limitations of execution-free assessment.

\textbf{Comparison with Existing Benchmarks.}
Table~\ref{tab:benchmark_comparison} summarizes the structural distinctions between current benchmarks. SWE-Cycle uniquely enforces the full issue resolution cycle while evaluating outputs through an execution-aware, fine-grained judge.

\begin{table*}[t]
\centering
\caption{\textbf{Comparison between SWE-Cycle and current software engineering benchmarks.}
}
\resizebox{\linewidth}{!}{
\begin{tabular}{l l ccc c cc}
\toprule
\multirow{2}{*}{\textbf{Benchmark}} & \multirow{2}{*}{\textbf{Scenario}} & \multicolumn{3}{c}{\textbf{Task}} & \multirow{2}{*}{\textbf{End-to-End}} & \multicolumn{2}{c}{\textbf{Evaluation Protocol}} \\
\cmidrule(lr){3-5}  \cmidrule(lr){7-8}
& & \textbf{Env.} & \textbf{Impl.} & \textbf{TestGen.} &   & \textbf{Judger} & \textbf{Partial Scoring} \\
\midrule
SWE-bench \& Variants~\citep{jimenez2024swebench, chowdhury2024swebenchverified, deng2025swebenchpro, zan2025multiswebench} & Issue & \xmark & \cmark & \xmark & \xmark & UnitTest & \xmark \\
EnvBench~\citep{eliseeva2025envbench} & Env Setup & \cmark & \xmark & \xmark & \xmark & UnitTest & \xmark \\
TestEval~\citep{wang2025testevalbenchmarkinglargelanguage} & Existing Code & \xmark & \xmark & \cmark & \xmark & UnitTest & \xmark \\
DevBench~\citep{golnari2026devbenchrealisticdeveloperinformedbenchmark}  & Greenfield & \xmark & \cmark & \cmark & \xmark & UnitTest+LLM & \xmark \\
PRDBench~\citep{wang2025prdbench} & Greenfield & \xmark & \cmark & \xmark & \cmark & Agent & \cmark \\
NL2Repo~\citep{ding2026nl2repobenchlonghorizonrepositorygeneration} & Greenfield & \xmark & \xmark & \xmark & \cmark & UnitTest & \xmark \\
\midrule
\rowcolor{gray!10} \textbf{SWE-Cycle (Ours)} & Issue & \cmark & \cmark & \cmark & \cmark & Agent & \cmark \\
\bottomrule
\end{tabular}
}

\label{tab:benchmark_comparison}
\end{table*}


\section{Methodology}

\begin{figure}[t]
    \centering
    \includegraphics[width=\linewidth]{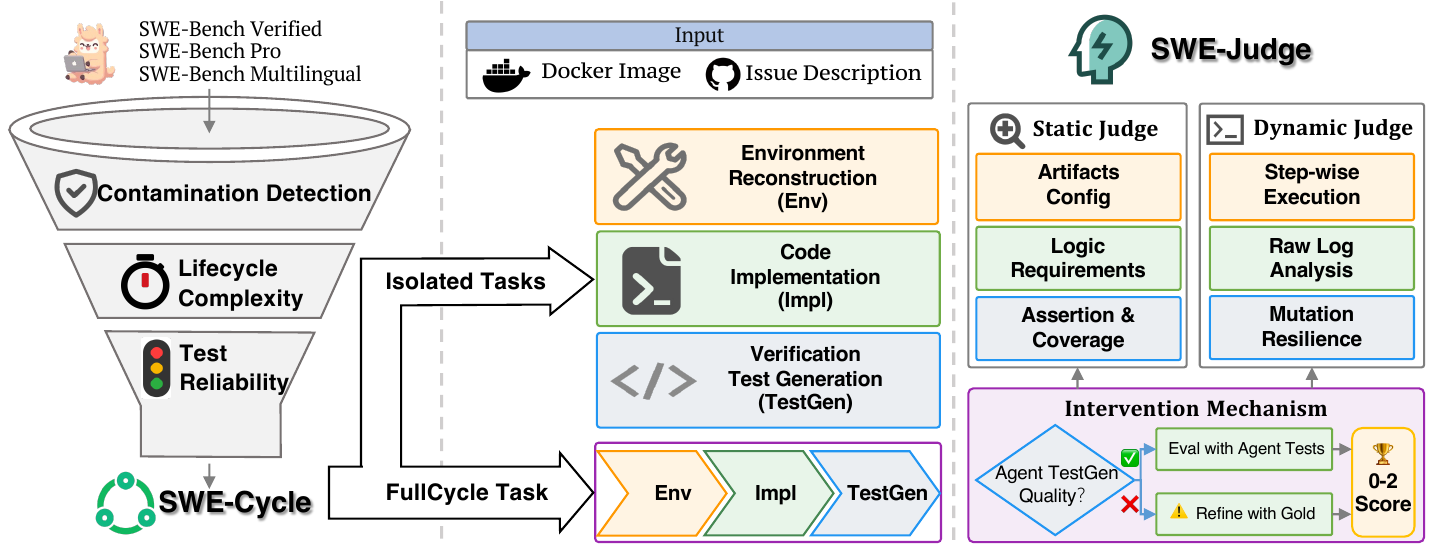}
    \caption{\textbf{Overview of the SWE-Cycle Framework.} \textit{Left:} High-quality instances are curated through a rigorous filtering pipeline. \textit{Center:} Agents execute environment reconstruction, code implementation, and test generation in either isolated tasks or the FullCycle task. \textit{Right:} {SWE-Judge} evaluates outputs via hybrid static-dynamic analysis and a test intervention mechanism to yield a robust 0-2 score.}
    \label{fig:main}
\end{figure}

To evaluate code agents across the software development lifecycle, we construct SWE-Cycle, a benchmark targeting the full issue resolution process. As detailed in Figure~\ref{fig:main}, the framework consists of three core components: dataset distillation to eliminate contamination, task formulation separating isolated capabilities from end-to-end execution, and hybrid evaluation via SWE-Judge. By combining static analysis with dynamic execution, this architecture overcomes the brittleness of predefined scripts, ensuring robust evaluation of autonomous workflows.

\subsection{Dataset Curation}
\label{subsec:dataset}
To evaluate code agents on issue resolution tasks, we source 1,531 initial instances from three established datasets: SWE-bench Verified~\citep{chowdhury2024swebenchverified}, SWE-bench Pro~\citep{deng2025swebenchpro}, and SWE-bench Multilingual~\citep{zan2025multiswebench}. However, the raw instances across these benchmarks suffer from data contamination, trivial task complexity, and invalid tests. To ensure the high quality required for assessing the complete issue resolution lifecycle, we design a three-stage filtering pipeline that resolves these flaws, distilling the initial pool into 489 rigorous instances.

\textbf{Contamination Detection.} Recent evidence confirms severe memorization in existing benchmarks: models achieve 35\% exact 5-gram match rates on reference patches and locate buggy files with 76\% accuracy without accessing the repository structure~\cite{liang2025swebenchillusionstateoftheartllms}. To eliminate training data leakage, we introduce a zero-context probing mechanism. Specifically, we prompt a recent LLM to generate a patch without providing any issue description. If the model successfully generates the correct patch without context, we assume the instance was seen during training. This step removes 128 contaminated instances.

\textbf{Lifecycle Complexity Filtering.} A benchmark targeting the complete development lifecycle requires tasks that reflect realistic engineering effort rather than instantaneous, isolated edits. To exclude trivial code tweaks, we filter instances using two criteria: (1) the pull request must contain at least one code review comment, and (2) the resolution cycle must span at least one day. These criteria remove same-day quick fixes and unreviewed submissions, ensuring the remaining instances represent complex maintenance tasks. After this step, the data pool is reduced to 523 instances.

\textbf{Test Reliability Filtering.} Recent audits reveal severe defects in existing benchmark test suites: nearly 60\% of SWE-bench Verified tests are flawed (often enforcing excessively narrow implementation details)~\citep{openai2026swebench}, and insufficient test coverage causes up to 28.4\% of incorrect patches to be falsely accepted~\citep{yu2025utboost}. To prevent these unreliable tests and environment dependency rot from compromising our evaluation, we execute the gold tests for each instance to verify strict state transitions: \texttt{FAIL\_TO\_PASS} tests must fail and \texttt{PASS\_TO\_PASS} tests must pass in the buggy state, and both must pass in the fixed state. This verification protocol removes 34 instances with corrupted test behavior.

Ultimately, SWE-Cycle yields 489 instances: 225 from SWE-bench Verified, 203 from SWE-bench Pro, and 61 from SWE-bench Multilingual. The detailed results of each filtering stage are summarized in Appendix~\ref{app:filter_pipeline}.

\subsection{Task Formulation}

In SWE-Cycle, we decompose the issue resolution lifecycle into three isolated tasks: 
\begin{itemize}[leftmargin=10pt]
    \item \textbf{Environment Reconstruction (Env)} task evaluates the environment configuration capabilities of code agents. Given only the source code, code agents must independently build the  execution environment and resolve all dependencies within a Docker container.
    \item \textbf{Code Implementation (Impl)} task assesses issue-driven development capabilities. Given an issue description and a preconfigured codebase, agents must modify the repository to implement the requested changes.
    \item \textbf{Verification Test Generation (TestGen)} task measures the capability to design effective unit tests. Given an issue description and a patched codebase, agents must generate tests targeting the specified issue.
\end{itemize}
We also integrate these phases into an end-to-end task, \textbf{FullCycle}, to simulate a realistic developer workflow. Given an issue description and a raw codebase, the agent must handle environment setup, code implementation, and test generation within a single autonomous session. This dual design allows us to evaluate specific engineering skills in isolation, while still testing the model's ability to manage the complete lifecycle without step-by-step guidance.

\subsection{SWE-Judge}

We introduce SWE-Judge to score the aforementioned tasks by combining static code review with dynamic execution. Unlike traditional script-driven methods, SWE-Judge applies task-specific criteria to yield independent static and dynamic scores (ranging from 0 to 2) for each task. The evaluation pipeline for each task is outlined below. Detailed scoring rubrics are available in Appendix~\ref{app:scoring_rubric}.

\textbf{Env Task Evaluation.} SWE-Judge first statically examines configuration artifacts for missing dependencies and version conflicts. It then executes a three-stage dynamic validation: the activation stage verifies toolchain setup, the import stage checks core package compilation, and the collection stage ensures the test runner successfully gathers cases without dependency failures. By pairing static artifact inspection with this step-by-step execution, SWE-Judge accurately captures the configuration status and isolates the exact step where a setup failure occurs.

\textbf{Impl Task Evaluation.} Traditional script evaluations for issue code implementation struggle to detect logical errors outside the predefined test coverage. SWE-Judge first extracts the core requirements, including the root cause, expected behavior, and critical edge cases, from the issue description and the official patch. It then statically reviews the submitted code against these requirements to verify its logical correctness. During dynamic execution, instead of relying solely on the rigid parsing of test outcomes, SWE-Judge analyzes the raw execution logs to accurately diagnose the actual runtime behavior. Consequently, SWE-Judge uncovers deep logical flaws that superficially pass rigid script evaluations, ultimately providing a more comprehensive and in-depth assessment.

\textbf{TestGen Task Evaluation.} Conventional script-driven evaluations merely require tests to fail on the buggy repository and pass on the patched one. This condition is easily exploited and ignores test quality metrics like assertion accuracy and scenario coverage. To address this, SWE-Judge supplements dynamic execution with static analysis: it evaluates assertion quality against the official test suite and maps scenario coverage to verify critical execution paths. This dual verification ensures the generated tests genuinely isolate the target bug rather than exploit evaluation loopholes.

\begin{algorithm}[t]
\caption{SWE-Judge Evaluation Pipeline for the FullCycle Task}
\label{alg:fullcycle_eval}
\begin{algorithmic}[1]
\Require Agent submissions $A_{\text{env}}, A_{\text{test}}, A_{\text{impl}}$; Official tests $T_{\text{gold}}$
\Ensure Scores for each task $(S_{\text{env}}, S_{\text{test}}, S_{\text{impl}})$, including dynamic and static scores

\State $S_{\text{env}}^{\text{stat}}, S_{\text{test}}^{\text{stat}}, S_{\text{impl}}^{\text{stat}} \leftarrow \textsc{EvalStatic}(A_{\text{env}}, A_{\text{test}}, A_{\text{impl}})$ \Comment{Static scores are always preserved}
\vspace{0.3em}

\State $S_{\text{env}}^{\text{dyn}} \leftarrow \textsc{EvalEnvDynamic}(A_{\text{env}})$
\State $S_{\text{env}} \leftarrow (S_{\text{env}}^{\text{stat}}, S_{\text{env}}^{\text{dyn}})$
\If{$S_{\text{env}}^{\text{dyn}} = 0$} \Comment{Halt: Upstream failure blocks execution}
    \State \Return $\big(S_{\text{env}}, (S_{\text{test}}^{\text{stat}}, 0), (S_{\text{impl}}^{\text{stat}}, 0)\big)$ 
\EndIf
\vspace{0.3em}

\State $S_{\text{test}}^{\text{dyn}} \leftarrow \textsc{EvalTestDynamic}(A_{\text{test}})$
\State $S_{\text{test}} \leftarrow (S_{\text{test}}^{\text{stat}}, S_{\text{test}}^{\text{dyn}})$
\If{\textsc{IsPoorQuality}($S_{\text{test}}$)} \Comment{Intervention: Refine agent tests}
    \State $T_{\text{exec}} \leftarrow \textsc{RefineTests}(A_{\text{test}}, T_{\text{gold}})$
\Else
    \State $T_{\text{exec}} \leftarrow A_{\text{test}}$
\EndIf
\vspace{0.3em}

\State $S_{\text{impl}}^{\text{dyn}} \leftarrow \textsc{EvalImplDynamic}(A_{\text{impl}}, T_{\text{exec}})$
\State $S_{\text{impl}} \leftarrow (S_{\text{impl}}^{\text{stat}}, S_{\text{impl}}^{\text{dyn}})$
\State \Return $(S_{\text{env}}, S_{\text{test}}, S_{\text{impl}})$
\end{algorithmic}
\end{algorithm}

\textbf{FullCycle Task Evaluation.} Traditional script evaluations rely entirely on predefined test suites, which is incompatible with the open-ended nature of the FullCycle task. As outlined in Algorithm \ref{alg:fullcycle_eval}, SWE-Judge addresses this by integrating the evaluation protocols from the isolated tasks into a sequential, fault-tolerant pipeline. SWE-Judge first conducts the Env evaluation, assigning zero to subsequent dynamic scores if the setup fails. Next, it evaluates the agent-generated unit tests. Since poor-quality generated tests cannot reliably verify the agent's subsequent code implementation, SWE-Judge intervenes: if the generated tests fail the evaluation, it refines the agent-generated tests based on the official gold tests to ensure accurate verification. Finally, SWE-Judge evaluates the code implementation by statically reviewing the submission and dynamically executing it against the finalized unit tests (the originally submitted tests or their refined versions). Ultimately, this step-wise, fault-tolerant approach ensures that upstream failures do not silently invalidate downstream assessments, enabling a robust evaluation across the complete issue resolution cycle.


\section{Experiments}
\label{sec:experiments}

In this section, we evaluate code agents powered by six state-of-the-art LLMs on SWE-Cycle to quantify their capabilities across the full issue-resolution lifecycle and validate the reliability of SWE-Judge. Specifically, our experiments, including an ablation study of SWE-Judge, are structured around four core research questions:

\begin{itemize}[leftmargin=10pt]
    \item \textbf{RQ1:} How reliable is the evaluation produced by SWE-Judge?
    \item \textbf{RQ2:} How do evaluated code agents perform across the four SWE-Cycle tasks?
    \item \textbf{RQ3:} How does SWE-Judge overcome the limitations of traditional script-based evaluations?
    \item \textbf{RQ4:} How do agent behaviors differ when resolving an issue end-to-end versus step-by-step?
\end{itemize}

\subsection{Experimental Setup}\label{sec:setup}
We evaluate code agents powered by six state-of-the-art LLMs spanning both proprietary and open-weight families: GPT-5.4~\citep{gpt54}, Claude-Sonnet-4.6~\citep{anthropic2025claude46}, Qwen-3.5~\citep{qwen35}, GLM-5.1~\citep{glm5team2026glm5}, Kimi-K2.5~\citep{kimi}, and MiniMax-M2.7~\citep{minimax27}. To guarantee reproducibility and support custom configurations, we adopt the open-source OpenCode\footnote{\url{https://github.com/anomalyco/opencode/releases/tag/v1.4.6}} framework across the evaluation pipeline. Each task executes within an isolated Docker container. We allocate 90 minutes per instance for the isolated tasks (Env, Impl, TestGen) and 3 hours for the FullCycle task; we evaluate all generated artifacts even if a timeout occurs. To balance assessment depth with computational cost, we use Claude-Opus-4.5~\citep{claude45} as the SWE-Judge backbone (see Appendix~\ref{appendix:eval-ablation} for robustness validation of backbone evaluation model). The evaluation agent runs directly in the same container to inherit the solving agent's environment and artifacts. 

\textbf{Metrics.} We measure agent performance in SWE-Cycle using four metrics. \textit{Static} assesses structural correctness or static analysis results without execution. \textit{Dynamic} (Dyn.) measures functional correctness through actual execution. Both are initially scored as 0, 1, or 2, then normalized to a 0--1 scale. \textit{Score} averages the \textit{Static} and \textit{Dynamic} results; for the FullCycle setting, this is macro-averaged across all three phases. Finally, \textit{Solve} denotes the perfect resolution rate, representing the fraction of instances where \textit{Score} equals 1. All metrics are reported as percentages.

\subsection{Reliability of SWE-Judge (RQ1)}\label{sec:rq1}

\begin{wraptable}{r}{0.4\linewidth}
\vspace{-10pt}
\small
\setlength{\tabcolsep}{9pt}
\centering
\caption{Alignment between human annotations and SWE-Judge across the four tasks. $N$ denotes the number of sampled instances.}\label{tab:rq1-alignment}
\begin{tabular}{lcc}
\toprule
\textbf{Task} & \textbf{N} & \textbf{Alignment\%} \\
\midrule
Env      & 143 & 99.3 \\
Impl     & 113 & 95.6 \\
TestGen  & 201 & 99.5 \\
FullCycle& 489 & 96.9 \\
\bottomrule
\end{tabular}

\end{wraptable}
To validate the reliability of SWE-Judge, we manually annotate samples  across all four tasks. For each task, we randomly select over 100 submissions generated by different agents across various issues. For the FullCycle task, we annotate one submission per issue.

As shown in Table~\ref{tab:rq1-alignment}, SWE-Judge aligns with human judgment in over 95\% of cases across all tasks. This confirms that SWE-Judge can reliably evaluate both isolated tasks and open-ended FullCycle resolutions without script guidance.  Appendix~\ref{app:annotation_protocol} details the annotation protocol and reports per-task alignment rates.

\subsection{Main Results (RQ2)}\label{sec:rq2}

\textbf{Isolated Tasks.} Table~\ref{tab:isolated-main} reports performance on the three isolated tasks, where gold inputs prevent cross-phase error propagation. Among these tasks, {Env} proves the most straightforward with solve rates reaching 78.1\%, whereas {Impl} remains the primary bottleneck for all models. Additionally, the noticeable drop from average scores to strict solve rates indicates that static and dynamic evaluations capture distinct errors, exposing multiple categories of LLM failures.

\textbf{End-to-End Task.} Table~\ref{tab:fullcycle-main} presents results for the end-to-end FullCycle task, revealing three key findings. (1) Compared to the isolated tasks, phase-specific scores show noticeable improvement. This indicates that engaging in interconnected phases provides agents with valuable context. For instance, the process of writing test cases directly enhances performance on the core code implementation. (2) Static scores are consistently lower than dynamic scores. This discrepancy occurs because agents often generate submissions that hack the corresponding tests to pass runtime execution, but these flawed implementations are caught by static analysis. (3) No model achieves a strict overall solve rate above 14\%, underscoring that completing the entire issue resolution lifecycle autonomously remains highly challenging for current code agents. Appendix~\ref{app:rq2_per_dataset} decomposes performance by dataset (Verified, Multi, Pro), and Appendix~\ref{app:rq2_efficiency} reports efficiency metrics including median token consumption and execution time per model.

\begin{wrapfigure}{hrtbp}{0.48\textwidth}
\vspace{-10pt}
    \centering
    \includegraphics[width=0.48\textwidth]{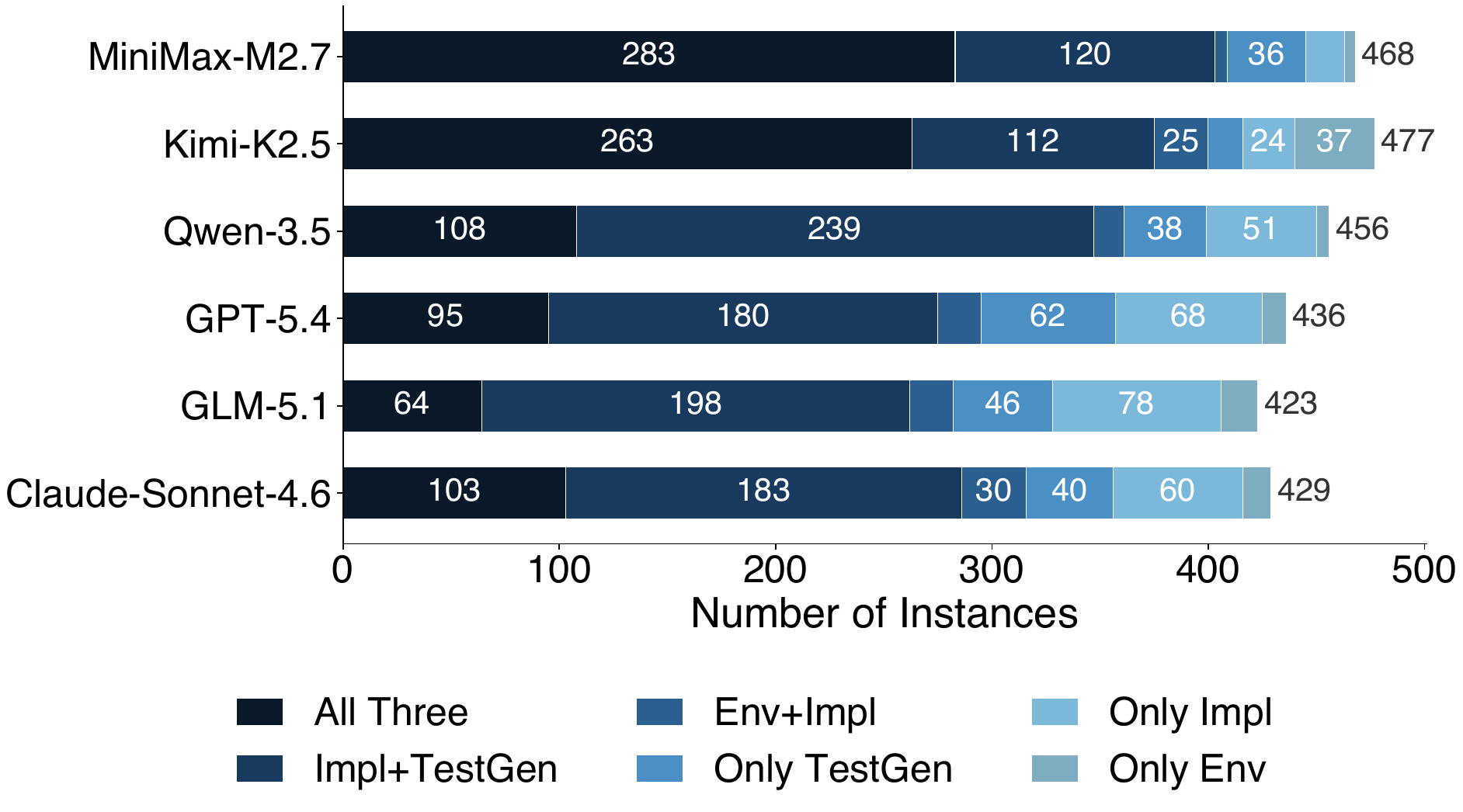}
    \caption{Distribution of failure categories across models in the FullCycle task.}
    \label{fig:failure_classification}
\end{wrapfigure} 

To further analyze the performance of code agents in the FullCycle task, we categorize the failures of unsuccessful instances in Figure~\ref{fig:failure_classification}. The distribution reveals that compound errors dominate across all evaluated models. Specifically, simultaneous failures in both implementation and test generation (Impl+TestGen), or across all three phases, constitute the vast majority of unsuccessful instances. In contrast, isolated single-phase failures account for a significantly smaller fraction of the total. This pattern indicates a strong cascading effect within the issue resolution lifecycle, where a breakdown in one component is highly correlated with failures in the interconnected phases.

\subsection{Effectiveness of SWE-Judge (RQ3)}\label{sec:rq3}
\begin{table}[t]
\centering
\caption{Leaderboard on isolated tasks. Best results are in \textbf{bold}, and second best are \underline{underlined}.}
\label{tab:isolated-main}
\small
\setlength{\tabcolsep}{2.5pt}
\renewcommand{\arraystretch}{1.0} 
\resizebox{\textwidth}{!}{%
\begin{tabular*}{\textwidth}{@{\extracolsep{\fill}}lcccccccccccc@{}}
\toprule
\multirow{2}{*}{\textbf{Model}} & \multicolumn{4}{c}{\textbf{Env}} & \multicolumn{4}{c}{\textbf{Impl}} & \multicolumn{4}{c}{\textbf{TestGen}} \\
\cmidrule(lr){2-5} \cmidrule(lr){6-9} \cmidrule(lr){10-13}
& \textbf{Static} & \textbf{Dyn.} & \textbf{Score} & \textbf{Solve} & \textbf{Static} & \textbf{Dyn.} & \textbf{Score} & \textbf{Solve} & \textbf{Static} & \textbf{Dyn.} & \textbf{Score} & \textbf{Solve} \\
\midrule
Claude-Sonnet-4.6 & \textbf{90.59} & \textbf{84.97} & \textbf{87.78} & \textbf{78.12} & \underline{68.00} & \textbf{57.16} & \textbf{62.58} & \textbf{40.08} & \textbf{91.00} & \textbf{85.79} & \textbf{88.39} & \textbf{67.28} \\
GLM-5.1 & 87.73 & \underline{76.48} & \underline{82.11} & \underline{73.01} & 65.95 & \underline{53.68} & 59.82 & 37.83 & \underline{85.89} & \underline{82.82} & \underline{84.36} & \underline{60.53} \\
GPT-5.4 & \underline{87.83} & 74.44 & 81.13 & 71.78 & \textbf{69.12} & 51.84 & \underline{60.48} & \underline{39.67} & 74.44 & 76.38 & 75.41 & 42.13 \\
Kimi-K2.5 & 81.09 & 68.80 & 74.95 & 61.34 & 64.38 & 51.06 & 57.72 & 34.46 & 80.79 & 82.44 & 81.61 & 52.07 \\
MiniMax-M2.7 & 56.13 & 60.94 & 58.54 & 45.40 & 52.86 & 43.46 & 48.16 & 30.88 & 52.66 & 55.11 & 53.89 & 33.33 \\
Qwen-3.5 & 82.62 & 75.05 & 78.83 & 66.46 & 56.24 & 43.35 & 49.80 & 28.83 & 72.90 & 76.89 & 74.90 & 46.01 \\
\bottomrule
\end{tabular*}}
\end{table}

\begin{table}[t]
\centering
\caption{Leaderboard on FullCycle task. Best results are in \textbf{bold}, and second best are \underline{underlined}.}
\label{tab:fullcycle-main}
\small
\setlength{\tabcolsep}{2.5pt}
\renewcommand{\arraystretch}{1.0} 
\resizebox{\textwidth}{!}{%
\begin{tabular*}{\textwidth}{@{\extracolsep{\fill}}lcccccccc@{}}
\toprule
\multirow{2}{*}{\textbf{Model}} & \multicolumn{2}{c}{\textbf{Env}} & \multicolumn{2}{c}{\textbf{Impl}} & \multicolumn{2}{c}{\textbf{TestGen}} & \multirow{2}{*}{\textbf{Score}} & \multirow{2}{*}{\textbf{Solve}} \\
\cmidrule(lr){2-3} \cmidrule(lr){4-5} \cmidrule(lr){6-7}
& \textbf{Static} & \textbf{Dyn.} & \textbf{Static} & \textbf{Dyn.} & \textbf{Static} & \textbf{Dyn.} & & \\
\midrule
Claude-Sonnet-4.6 & 84.56 & \textbf{97.55} & 61.35 & \textbf{89.67} & \underline{68.61} & 81.39 & \underline{80.52} & \underline{12.27} \\
GLM-5.1 & \textbf{87.93} & \underline{96.73} & \underline{61.76} & \underline{89.37} & \textbf{70.35} & \textbf{82.82} & \textbf{81.49} & \textbf{13.50} \\
GPT-5.4 & \underline{87.22} & 93.15 & \textbf{62.88} & 88.96 & 65.95 & \underline{81.49} & 79.94 & 10.84 \\
Kimi-K2.5 & 65.88 & 89.81 & 58.91 & 80.90 & 56.22 & 71.35 & 70.51 & 2.15 \\
MiniMax-M2.7 & 46.01 & 55.83 & 31.70 & 43.35 & 26.38 & 35.79 & 39.84 & 4.29 \\
Qwen-3.5 & 85.07 & 94.89 & 56.24 & 80.06 & 56.34 & 74.03 & 74.44 & 6.75 \\
\bottomrule
\end{tabular*}}
\end{table}

\textbf{SWE-Judge vs. Script Evaluation.} To evaluate the effectiveness of SWE-Judge, we sampled 371 instances across all three phases where it diverged from traditional script metrics. As shown in Table~\ref{tab:rq3-disagree}, human adjudication confirms that SWE-Judge is correct in 98.6\% of these disagreements, achieving 100\% accuracy in the TestGen and Env phases, whereas scripts were correct in only 0.5\% of cases. Analyzing these script failures reveals three primary structural flaws: excessive strictness (36.0\%) that penalizes functionally equivalent alternatives, evaluation breakdown (32.8\%) caused by brittle static pipelines, and excessive leniency (27.0\%) that allows superficial or overfitted fixes to bypass basic execution checks. By semantically interpreting test outputs and dynamically adapting to the execution context, SWE-Judge resolves these systemic biases. Detailed category results and case studies are presented in Appendix~\ref{app:disagree_cases}.

\begin{wraptable}{r}{0.48\linewidth}
\vspace{-10pt}
\setlength{\tabcolsep}{4pt}
\small
\centering
\caption{Human adjudication of 371 disagreements between SWE-Judge and script-based metrics. Percentages denote how often humans agreed with each method.}\label{tab:rq3-disagree}
\begin{tabular}{lcccc}
\toprule
\textbf{Task} & \textbf{N} & \textbf{SWE-Judge} & \textbf{Script} & \textbf{Neither} \\
\midrule
Impl    & 93  & 94.6\% & 2.2\% & 3.2\% \\
TestGen & 173 & 100.0\% & 0.0\% & 0.0\% \\
Env     & 105 & 100.0\% & 0.0\% & 0.0\% \\
\midrule
Overall & 371 & 98.6\% & 0.5\% & 0.8\% \\
\bottomrule
\end{tabular}
\vspace{-10pt}
\end{wraptable}



\textbf{Adaptive Evaluation Workflow.} To understand how SWE-Judge operates without gold tests, we analyzed its trajectories across all valid FullCycle instances. Table~\ref{tab:rq3-behavior} reports the average invocations per trajectory (Mean) and the percentage of evaluations using each action (Coverage). Instead of relying on static heuristics, SWE-Judge executes a systematic pipeline. It universally anchors on code diffs and reference comparisons, then selectively runs tests and build verifications to avoid false positives from broken environments. Crucially, it compensates for missing or flawed agent artifacts by actively writing custom evaluation scripts (34.6\%) and using fault injection to verify test robustness (4.8\%). These dynamic behaviors prove SWE-Judge's capability for end-to-end evaluation. Appendix~\ref{app:judge_workflow} presents three representative workflows. Appendix~\ref{app:appendix_blind} validates our design choice to anchor on reference patches: removing them causes a severe 18.4 percentage point inflation in static scores, further highlighting the necessity of SWE-Judge's comprehensive dynamic checks to ensure accurate evaluation.



\subsection{End-to-End vs.\ Isolated Evaluation (RQ4)}\label{sec:rq4}

\begin{figure*}[t]
\centering
\begin{subfigure}[t]{0.48\textwidth}
  \centering
  \includegraphics[width=\textwidth]{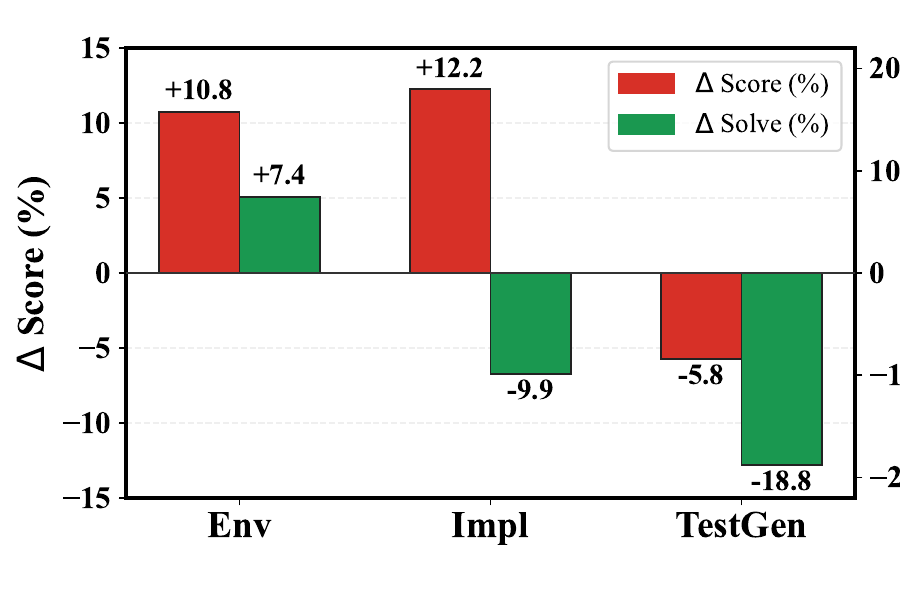}
  \vspace{-8pt}
  \caption{}
  \label{fig:rq4-shift}
\end{subfigure}
\hfill
\begin{subfigure}[t]{0.48\textwidth}
  \centering
  \includegraphics[width=\textwidth]{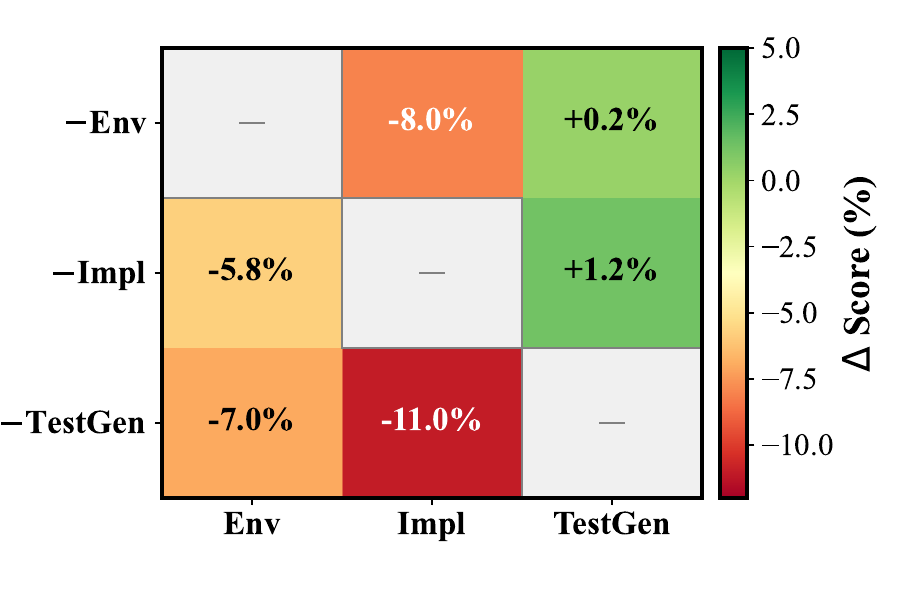}
  \caption{}
  \label{fig:rq4-ablation}
\end{subfigure}
\caption{End-to-end integration effects. (a) Per-dimension score and solve rate change ($\Delta = \text{FullCycle} - \text{Isolated}$) across three tasks. (b) Score degradation of remaining phases when one phase is removed from the FullCycle task.}
\label{fig:rq4-combined}
\end{figure*}

To quantify how end-to-end integration reshapes model behavior, we evaluate 489 instances paired across FullCycle and isolated tasks. Figure~\ref{fig:rq4-shift} visualizes the performance shift by plotting the absolute difference ($\Delta = \text{FullCycle} - \text{Isolated}$) in average score and overall solve rate across the three tasks. The results show a straightforward trend: integration improves upstream performance but degrades downstream metrics. Upstream Env benefits for most models because subsequent runtime execution exposes configuration defects that models then return to fix. Midstream Impl gains average dynamic functionality through this iterative write-run-fix loop, but the continuous patching compromises static structural quality, dropping the solve rate. Downstream TestGen degrades across all metrics. This occurs because models often hack the verification step by writing trivial tests that simply pass their own implementations, aiming to terminate the task as quickly as possible rather than rigorously testing the code.

\begin{wraptable}{r}{0.48\linewidth}
\vspace{-6pt}
\centering
\caption{SWE-Judge action distribution across FullCycle evaluations.}\label{tab:rq3-behavior}
\resizebox{\linewidth}{!}{
\begin{tabular}{lcc}
\toprule
\textbf{Action} & \textbf{\# Avg} & \textbf{Coverage} \\
\midrule
Code Review (git diff)      & 9.7  & 100\% \\
Reference Comparison (Read) & 7.9  & 100\% \\
Env \& File Inspection      & 16.5 & 99.9\% \\
Test Execution              & 1.6  & 45.7\% \\
Build verification          & 0.4  & 36.1\% \\
Adaptive eval scripting     & ---  & 34.6\% \\
Fault injection             & ---  & 4.8\% \\
\midrule
Total steps                 & 39.5 & --- \\
\bottomrule
\end{tabular}
}
\vspace{-5pt}
\end{wraptable}

To measure the inter-dependency of these tasks, Figure~\ref{fig:rq4-ablation} maps a phase-ablation experiment. We systematically remove a single phase from the FullCycle setup and record the resulting score degradations across the remaining active phases. These results confirm that the phases are tightly interlocked. Removing any single phase causes scores in the others to drop. Notably, removing the downstream test phase severely penalizes upstream environment and code performance. This highlights that a code agent's verification capability is a critical driver of overall success. End-to-end evaluation measures an orchestration capability that isolated tasks structurally fail to capture (a detailed analysis is provided in Appendix~\ref{app:rq4}).

\section{Conclusions \& Limitations}
\label{sec:discussion}
In this paper, we propose SWE-Cycle, an innovative full-lifecycle issue resolution benchmark specifically designed to address critical challenges in evaluating the end-to-end autonomy of code agents.
Alongside this benchmark, we introduce SWE-Judge, a robust agentic evaluator that uniquely integrates static and dynamic analysis, thereby providing comprehensive and reliable assessments.
Our rigorous human validation and case studies demonstrate that SWE-Judge effectively overcomes the brittleness inherent in traditional evaluation scripts, enabling accurate and scalable evaluation across complex software engineering tasks.
Additionally, leveraging this comprehensive framework, we uncover a critical blind spot in current LLM evaluation: optimizing localized accuracy in isolated tasks does not yield full-lifecycle autonomy.
Results confirm that during end-to-end integration, cross-phase feedback paradoxically boosts dynamic correctness while accumulating structural debt, which ultimately degrades downstream test generation.
A current limitation of SWE-Cycle lies in its evaluation mechanism. To ensure rigorous evaluation accuracy, our current evaluator relies on the Claude API, which inevitably makes the framework susceptible to external API fluctuations and version updates. To address this issue, future work will focus on enhancing open-weight models to serve as robust verifiers, thereby providing a stable, reliable, and independent evaluation backend.
Overall, SWE-Cycle and SWE-Judge collectively represent a significant advancement in the evaluation of code agents. They necessitate a paradigm shift from isolated code generation to global planning and long-term maintainability, paving the way for the next evolution of autonomous software engineering by providing the rigorous standards required to guide future agent development.
%


\bibliographystyle{plainnat}
\bibliography{references}

\newpage
\appendix

\input{Appendix}



\end{document}

%% file: Appendix.tex

\input{Appendix/dataset_construction}
\input{Appendix/scoring_rubric}

\input{Appendix/setup-eval_model_ablation}

\input{Appendix/rq1_annotation}

\input{Appendix/rq2_per_dataset}
\input{Appendix/rq2_efficiency}

\input{Appendix/rq3_script_failures}
\input{Appendix/rq3_judge_workflow}
\input{Appendix/ablation_blind}

\input{Appendix/rq4_details}

\input{Appendix/broader_impacts}

%% file: Appendix/dataset_construction.tex
\section{Dataset Construction}\label{app:dataset}\label{app:filter_pipeline}

\subsection{Detailed Filtering Statistics}
Table~\ref{tab:filter_pipeline} reports the number of retained instances at each filtering stage. Below, we detail the concrete procedure for each stage. We run DeepSeek-V3.2~\citep{deepseekai2025deepseekv32pushingfrontieropen} for Contamination Detection.

\begin{table}[htbp]
\centering
\caption{The SWE-Cycle filtering pipeline and the number of retained instances at each step.}
\label{tab:filter_pipeline}
\begin{tabular}{lcccc}
\toprule
Targeted Problem & \textbf{Verified} & \textbf{Pro} & \textbf{Multi} & Total Retained \\ \midrule
Initial Source Pool & 500 & 731 & 300 & 1,531 \\ \midrule
Contamination Detection & 499 & 607 & 297 & 1,403 \\ \midrule
Lifecycle Complexity Filtering & 239 & 209 & 75 & 523 \\ \midrule
Test Reliability Filtering & 225 & 203 & 61 & \textbf{489} \\ \bottomrule
\end{tabular}
\end{table}

\subsection{Language Distribution  and Environment Setup}
The final 489 instances span 9 programming languages: Python (68.9\%), Go (18.0\%), C (3.3\%), Ruby (2.7\%), JavaScript (2.5\%), Rust (1.4\%), TypeScript (1.2\%), Java (1.2\%), and PHP (0.8\%).
For Impl and TestGen tasks, each instance includes a pre-built Docker image with pinned runtimes and dependencies to ensure reproducible evaluation. In contrast, the Env task starts from a minimal base image containing only the OS and language runtime; the agent must independently resolve all project-level dependencies, build configurations, and toolchain setup. 

%% file: Appendix/scoring_rubric.tex
\section{Scoring Rubric}\label{app:scoring_rubric}

This appendix details the scoring criteria used by SWE-Judge for each task type in SWE-Cycle.
All task types except FullCycle use a two-dimensional rubric (Static Analysis + Dynamic Execution), each scored 0--2, yielding a maximum of 4 points.
FullCycle uses three dimensions (Environment, Code, Test), each with a static and dynamic sub-dimension (0--2 each), yielding a maximum of 12 points.
The final metric is the \textbf{score ratio} = total score / maximum achievable score. Detailed eval prompts are in GitHub repo.


\subsection{Env Task}\label{app:rubric_environment}

This task evaluates whether the agent correctly configured the required project dependencies and test runtime environment.

\begin{table}[H]
\centering
\small
\caption{Scoring rubric for the Environment task (max 4 points).}
\label{tab:rubric_env}
\begin{tabular}{ccp{9.5cm}}
\toprule
\textbf{Dimension} & \textbf{Score} & \textbf{Criteria} \\ \midrule
\multirow{3}{*}{\shortstack{Static\\Analysis}}
  & 0 & No environment configured: core dependencies completely absent; \texttt{setup.sh} missing or fundamentally broken. \\
  & 1 & Environment created, core dependencies basically installed, but obvious shortcomings: missing dependencies, version incompatibility risk, \texttt{setup.sh} not reproducible. \\
  & 2 & Environment complete: all necessary dependencies correctly installed, \texttt{setup.sh} clear and reproducible, no compatibility issues. \\ \midrule
\multirow{3}{*}{\shortstack{Dynamic\\Execution}}
  & 0 & Cannot activate environment; or all tests fail after activation. \\
  & 1 & Can activate, but tests partially fail (some FAIL\_TO\_PASS not passing or PASS\_TO\_PASS regressions). \\
  & 2 & Can activate, all FAIL\_TO\_PASS pass, PASS\_TO\_PASS has no new failures. \\
\bottomrule
\end{tabular}
\end{table}

\subsection{Impl Task}\label{app:rubric_development}

This task evaluates whether the agent's code patch correctly and completely resolves the reported bug.

\begin{table}[H]
\centering
\small
\caption{Scoring rubric for the Development task (max 4 points).}
\label{tab:rubric_dev}
\begin{tabular}{ccp{9.5cm}}
\toprule
\textbf{Dimension} & \textbf{Score} & \textbf{Criteria} \\ \midrule
\multirow{3}{*}{\shortstack{Static Code\\Analysis}}
  & 0 & Patch is empty or unrelated to the problem; or contains a fundamental logic error. \\
  & 1 & Fix direction is correct but has obvious shortcomings: incomplete logic, or change scope exceeds what is necessary. \\
  & 2 & Fix logic is correct and complete, change scope is reasonable, targets the root cause with no significant side effects. \\ \midrule
\multirow{3}{*}{\shortstack{Dynamic\\Execution}}
  & 0 & Tests cannot run; or FAIL\_TO\_PASS tests have failures. \\
  & 1 & All FAIL\_TO\_PASS tests pass, but PASS\_TO\_PASS has new failures (regression). \\
  & 2 & All FAIL\_TO\_PASS tests pass, PASS\_TO\_PASS has no new failures. \\
\bottomrule
\end{tabular}
\end{table}

\subsection{TestGen Task}\label{app:rubric_testcase}

This task evaluates the quality, coverage, and effectiveness of the test cases written by the agent. 
The dynamic dimension uses a two-phase protocol: Phase~1 reverts the code fix (expect tests to \textbf{fail}); Phase~2 restores the fix (expect tests to \textbf{pass}).

\begin{table}[H]
\centering
\small
\caption{Scoring rubric for the TestCase task (max 4 points).}
\label{tab:rubric_testcase}
\begin{tabular}{ccp{9.5cm}}
\toprule
\textbf{Dimension} & \textbf{Score} & \textbf{Criteria} \\ \midrule
\multirow{3}{*}{\shortstack{Static\\Analysis}}
  & 0 & \texttt{eval.sh} missing; contains git operations; all meaningless assertions (\texttt{assert True}); content unrelated to the bug. \\
  & 1 & Direction correct but quality insufficient: imprecise assertions, single scenario, low alignment with gold test scenarios; or tests merely mirror the code diff without additional insight. \\
  & 2 & Good design: targets bug core behavior, meaningful assertions, covers main scenarios with good alignment to gold test coverage; demonstrates understanding beyond the diff (e.g., edge cases, boundary conditions). \\ \midrule
\multirow{3}{*}{\shortstack{Two-Phase\\Dynamic}}
  & 0 & Phase~1 passes on buggy code (cannot detect bug); or \texttt{eval.sh} cannot execute. \\
  & 1 & Phase~1 fails but with imprecise cause (import error / setup failure / permanently-false assertion); or Phase~2 still fails. \\
  & 2 & Phase~1 fails precisely (points to the bug), Phase~2 passes. \\
\bottomrule
\end{tabular}
\end{table}


\subsection{FullCycle Task}\label{app:rubric_fullcycle}
This task evaluates the agent's end-to-end performance. The scoring criteria aggregate the individual rubrics from the Environment, TestCase, and Development tasks, with execution dependencies following the pipeline pseudocode detailed in the main text.
CODE\_DYNAMIC uses the agent's tests directly if \textsc{IsPoorQuality}($S_{\text{test}}$) returns false (Algorithm~\ref{alg:fullcycle_eval}); otherwise the evaluator refines the agent-generated tests based on the official gold tests to ensure accurate verification.

\begin{table}[htbp]
\centering
\small
\caption{Scoring rubric for the FullCycle task (max 12 points).}
\label{tab:rubric_fullcycle}
\begin{tabular}{cccp{8cm}}
\toprule
\textbf{Dim.} & \textbf{Sub-dim.} & \textbf{Score} & \textbf{Criteria} \\ \midrule
\multirow{6}{*}{\rotatebox[origin=c]{90}{\textbf{ENV (0--4)}}}
  & \multirow{3}{*}{Static}
    & 0 & No environment created; core dependencies absent; \texttt{setup.sh} has fundamental errors. \\
  & & 1 & Environment created, basically installed, but obvious gaps (missing deps / version issues / no editable install). \\
  & & 2 & Complete and excellent: all deps correctly installed, \texttt{setup.sh} clear and reproducible. \\ \cmidrule{2-4}
  & \multirow{3}{*}{Dynamic}
    & 0 & Layer~1 fails: cannot activate environment / toolchain missing. \\
  & & 1 & Layer~1 OK, but Layer~2 or Layer~3 fails (import/build/collection errors). \\
  & & 2 & Layers~1--3 all pass. \\ \midrule
\multirow{6}{*}{\rotatebox[origin=c]{90}{\textbf{TEST (0--4)}}}
  & \multirow{3}{*}{Static}
    & 0 & No test code; only meaningless assertions; \texttt{eval.sh} contains git operations. \\
  & & 1 & Direction correct, quality insufficient: covers only one scenario, imprecise assertions, low alignment with gold scenarios. \\
  & & 2 & Good: targets bug core behavior, meaningful assertions, covers main scenarios, good alignment with gold test coverage. \\ \cmidrule{2-4}
  & \multirow{3}{*}{Dynamic}
    & 0 & Phase~1 passes on buggy code (cannot detect bug); or no test code exists. \\
  & & 1 & Phase~1 fails but with imprecise cause (ImportError/setup failure); or Phase~2 still fails. \\
  & & 2 & Phase~1 fails precisely (points to bug), Phase~2 passes. \\ \midrule
\multirow{6}{*}{\rotatebox[origin=c]{90}{\textbf{CODE (0--4)}}}
  & \multirow{3}{*}{Static}
    & 0 & No functional code changes; fundamental logic error; completely unrelated to the issue. \\
  & & 1 & Direction correct but incomplete: misses key edge cases, overly broad scope, or partial fix. \\
  & & 2 & Complete and precise: targets root cause, changes minimal, edge cases handled, highly aligned with gold direction. \\ \cmidrule{2-4}
  & \multirow{3}{*}{Dynamic}
    & 0 & Environment failure prevents running; or all functional tests fail. \\
  & & 1 & Some functional tests pass (core logic OK, missing edge/secondary scenarios). \\
  & & 2 & All (or almost all) tests pass. \\
\bottomrule
\end{tabular}
\end{table}

Detailed eval prompts are shown in \href{https://anonymous.4open.science/r/SWE-Cycle-462E/ccb_templates/fullpipe/eval_prompt.md.j2}{GitHub repo}.




%% file: Appendix/setup-eval_model_ablation.tex
\section{Eval Model Robustness}\label{appendix:eval-ablation}

We test whether SWE-Judge's scores depend on the choice of judge model by having Claude-Opus-4.5 and GPT-5.4 independently evaluate identical agent outputs. We randomly assign 414 instances across 6 coding models for all four task types. Every trial is scored by both eval models, using solve rate as the primary metric. We report solve rate as the primary metric.

\begin{table}[h]
\centering
\caption{Solve rate comparison between two eval models across task types. Diff = Claude-Opus-4.5 $-$ GPT-5.4.}
\label{tab:ablation-solve}
\begin{tabular}{lccc}
\toprule
\textbf{Task Type} & \textbf{Claude-Opus-4.5} & \textbf{GPT-5.4} & \textbf{Diff} \\
\midrule
Impl & 0.4192 & 0.4034 & +0.016 \\
TestGen & 0.3985 & 0.3718 & +0.027 \\
Env & 0.6032 & 0.5929 & +0.010 \\
FullCycle & 0.0630 & 0.0419 & +0.021 \\
\bottomrule
\end{tabular}
\end{table}

The solve rate difference between Claude-Opus-4.5 and GPT-5.4 is within 3\% across all four task types (largest: +2.7\% on TestGen). The relative ranking of coding models is preserved under both eval models.

\paragraph{Same-family bias check.}
Our default eval model (Claude-Opus-4.5) shares a model family with one coding model (Claude Sonnet 4.6). We compare the mean score boost from Claude-Opus-4.5 (per-trial Claude-Opus-4.5 $-$ GPT-5.4) for Claude Sonnet against the average boost for the other five coding models. Across all four task types, Claude Sonnet's boost is equal to or lower than the average of other models (Impl: $-$0.001 vs.\ +0.006; Env: +0.003 vs.\ +0.008; FullCycle: +0.143 vs.\ +0.194), except TestGen where the difference is negligible (+0.025 vs.\ +0.024). Claude-Opus-4.5 does not exhibit same-family scoring bias.

%% file: Appendix/rq1_annotation.tex
\section{Human Annotation for SWE-Judge Reliability}\label{app:annotation_protocol}

We annotate 946 instances across all four tasks to validate SWE-Judge's reliability.

\textbf{Scope.} For isolated tasks (Impl, TestGen, Env), we annotate 457 cases. For FullCycle, we annotate all 489 instances (100\% coverage).

\textbf{Annotation Criteria.} Annotators adjudicate the factual correctness of SWE-Judge's verdict by cross-referencing multiple evidence sources: the issue description, gold reference patch, agent's submitted patch, and execution logs. For isolated tasks, annotators determine whether SWE-Judge or the script evaluator is correct. For FullCycle, annotators independently assign scores on 6 dimensions (each 0--2) using the same rubric as SWE-Judge and flag cases of overrating, underrating, or hallucination.

\textbf{Quality Control.} A second researcher conducted a blind spot-check on 48 instances. The audit found 1 discrepancy (2.1\%), which was corrected.

\subsection{Human Annotators}
Our annotation team consisted of three full-time professionals, each holding at least a bachelor's degree in computer-related disciplines (e.g., information security and software engineering) and possessing over two years of Python development experience. They maintained an average annotation rate of one task per hour, with daily working hours capped at eight. All annotators were compensated in strict compliance with local labor regulations.

\subsection{Isolated Tasks Detailed Results}

Table~\ref{tab:app-sp-detail} reports the per-task alignment between human annotations and SWE-Judge for isolated tasks.

\begin{table}[htbp]
\centering
\caption{Per-task alignment on isolated tasks. Alignment = annotator confirms SWE-Judge is correct.}
\label{tab:app-sp-detail}
\begin{tabular}{lrc}
\toprule
\textbf{Task} & \textbf{N} & \textbf{Alignment\%} \\
\midrule
\textsc{Impl}    & 113 & 95.6\% \\
\textsc{TestGen} & 201 & 99.5\% \\
\textsc{Env}     & 143 & 99.3\% \\
\midrule
\textbf{Overall} & \textbf{457} & \textbf{98.5\%} \\
\bottomrule
\end{tabular}
\end{table}

For isolated tasks, our annotation covers all cases where SWE-Judge and the script evaluator disagree (371 instances), plus a 10\% random sample of agreement cases (86 instances). Among the agreement sample, 97.8\% (85/86) were confirmed correct by human annotators, indicating that evaluator consensus reliably reflects true correctness with negligible risk of systematic co-failure.

\subsection{\textsc{FullCycle} Bias Analysis}

Table~\ref{tab:app-fp-bias} reports the direction of scoring disagreements: how often SWE-Judge scores higher vs.\ lower than human annotators.

\begin{table}[htbp]
\centering
\caption{Bias direction per dimension: frequency of SWE-Judge scoring higher (overrate) or lower (underrate) than human.}
\label{tab:app-fp-bias}
\begin{tabular}{lcccc}
\toprule
\textbf{Dimension} & \textbf{N} & \textbf{Judge $>$ Human} & \textbf{Exact} & \textbf{Judge $<$ Human} \\
\midrule
ENV static    & 486 & 2 (0.4\%) & 484 (99.6\%) & 0 (0.0\%) \\
ENV dynamic   & 483 & 3 (0.6\%) & 479 (99.2\%) & 1 (0.2\%) \\
CODE static   & 485 & 6 (1.2\%) & 478 (98.6\%) & 1 (0.2\%) \\
CODE dynamic  & 482 & 0 (0.0\%) & 482 (100.0\%) & 0 (0.0\%) \\
TEST static   & 485 & 4 (0.8\%) & 478 (98.6\%) & 3 (0.6\%) \\
TEST dynamic  & 482 & 1 (0.2\%) & 476 (98.8\%) & 5 (1.0\%) \\
\midrule
\textbf{Total} & \textbf{2,903} & \textbf{16 (0.6\%)} & \textbf{2,877 (99.1\%)} & \textbf{10 (0.3\%)} \\
\bottomrule
\end{tabular}
\end{table}

SWE-Judge slightly favors overrating (16 cases) over underrating (10 cases), but both are below 1\%. TEST dynamic is the only dimension where underrating exceeds overrating (5 vs.\ 1). Note: per-dimension $N$ varies slightly below 489 because a small number of evaluation runs failed to produce a parseable score for the corresponding dimension.

%% file: Appendix/rq2_per_dataset.tex
\section{Per-Dataset Results}\label{app:rq2_per_dataset}

Tables~\ref{tab:rq2-app-dev}--\ref{tab:rq2-app-env} decompose the aggregate Isolated results (Table~\ref{tab:isolated-main}) by benchmark source. Static, Dynamic, and Score are percentages normalized to $[0, 100]$; Solve is the fraction of instances achieving a perfect score.

\begin{table}[htbp]
\centering
\caption{\textbf{Per-dataset CodeImpl performance.} Best results are in \textbf{bold}, and second best are \underline{underlined}.}
\label{tab:rq2-app-dev}
\small
\setlength{\tabcolsep}{2.5pt}
\renewcommand{\arraystretch}{1.0}
\resizebox{\textwidth}{!}{%
\begin{tabular*}{\textwidth}{@{\extracolsep{\fill}}lcccccccccccc@{}}
\toprule
\multirow{2}{*}{\textbf{Model}} & \multicolumn{4}{c}{\textbf{Verified}} & \multicolumn{4}{c}{\textbf{Multi.}} & \multicolumn{4}{c}{\textbf{Pro}} \\
\cmidrule(lr){2-5} \cmidrule(lr){6-9} \cmidrule(lr){10-13}
& \textbf{Static} & \textbf{Dyn.} & \textbf{Score} & \textbf{Solve} & \textbf{Static} & \textbf{Dyn.} & \textbf{Score} & \textbf{Solve} & \textbf{Static} & \textbf{Dyn.} & \textbf{Score} & \textbf{Solve} \\
\midrule
Claude-Sonnet-4.6 & \underline{74.00} & \textbf{70.00} & \textbf{72.00} & \textbf{53.3} & 60.50 & \textbf{55.50} & \textbf{58.00} & \textbf{34.4} & \underline{63.50} & \textbf{43.50} & \textbf{53.50} & 28.1 \\
GLM-5.1 & 71.50 & \underline{66.00} & 68.75 & 48.0 & \underline{60.50} & 51.50 & 56.00 & 29.5 & 61.50 & \underline{40.50} & 51.00 & \textbf{30.5} \\
GPT-5.4 & \textbf{75.50} & 64.50 & \underline{70.00} & \underline{52.0} & 58.00 & 50.00 & 54.00 & 31.1 & \textbf{65.50} & 38.00 & \underline{51.75} & \underline{29.1} \\
Kimi-K2.5 & 70.81 & 65.16 & 67.99 & 48.0 & 56.03 & 49.14 & 52.59 & 24.1 & 59.54 & 35.57 & 47.55 & 22.2 \\
MiniMax-M2.7 & 68.50 & 61.50 & 65.00 & 47.1 & 43.50 & 37.00 & 40.25 & 24.6 & 38.00 & 25.50 & 31.75 & 16.3 \\
Qwen-3.5 & 55.50 & 49.00 & 52.25 & 36.4 & \textbf{61.50} & \underline{51.50} & \underline{56.50} & \underline{32.8} & 55.50 & 34.50 & 45.00 & 19.2 \\
\bottomrule
\end{tabular*}}
\end{table}

\begin{table}[htbp]
\centering
\caption{\textbf{Per-dataset TestGen performance.} Best results are in \textbf{bold}, and second best are \underline{underlined}.}
\label{tab:rq2-app-tc}
\small
\setlength{\tabcolsep}{2.5pt}
\renewcommand{\arraystretch}{1.0}
\resizebox{\textwidth}{!}{%
\begin{tabular*}{\textwidth}{@{\extracolsep{\fill}}lcccccccccccc@{}}
\toprule
\multirow{2}{*}{\textbf{Model}} & \multicolumn{4}{c}{\textbf{Verified}} & \multicolumn{4}{c}{\textbf{Multi.}} & \multicolumn{4}{c}{\textbf{Pro}} \\
\cmidrule(lr){2-5} \cmidrule(lr){6-9} \cmidrule(lr){10-13}
& \textbf{Static} & \textbf{Dyn.} & \textbf{Score} & \textbf{Solve} & \textbf{Static} & \textbf{Dyn.} & \textbf{Score} & \textbf{Solve} & \textbf{Static} & \textbf{Dyn.} & \textbf{Score} & \textbf{Solve} \\
\midrule
Claude-Sonnet-4.6 & \textbf{92.50} & \textbf{96.50} & \textbf{94.50} & \textbf{85.8} & \textbf{84.50} & \underline{86.00} & \underline{85.25} & \textbf{70.5} & \textbf{91.00} & \textbf{73.50} & \textbf{82.25} & \textbf{45.8} \\
GLM-5.1 & \underline{88.00} & \underline{93.50} & \underline{90.75} & \underline{77.3} & \underline{83.50} & \textbf{90.00} & \textbf{86.75} & \underline{68.9} & \underline{84.50} & 68.50 & 76.50 & 39.4 \\
GPT-5.4 & 69.00 & 78.50 & 73.75 & 40.9 & 73.00 & 81.00 & 77.00 & 50.8 & 81.00 & \underline{72.50} & \underline{76.75} & \underline{40.9} \\
Kimi-K2.5 & 82.00 & 92.89 & 87.44 & 63.6 & 74.56 & 81.58 & 78.07 & 54.4 & 81.19 & 71.04 & 76.11 & 38.6 \\
MiniMax-M2.7 & 71.00 & 79.50 & 75.25 & 51.6 & 31.00 & 32.00 & 31.50 & 21.3 & 38.50 & 35.00 & 36.75 & 16.7 \\
Qwen-3.5 & 73.50 & 86.50 & 80.00 & 56.9 & 69.00 & 81.00 & 75.00 & 42.6 & 73.50 & 65.00 & 69.25 & 35.0 \\
\bottomrule
\end{tabular*}}
\end{table}

\begin{table}[htbp]
\centering
\caption{\textbf{Per-dataset Env performance.} Best results are in \textbf{bold}, and second best are \underline{underlined}.}
\label{tab:rq2-app-env}
\small
\setlength{\tabcolsep}{2.5pt}
\renewcommand{\arraystretch}{1.0}
\resizebox{\textwidth}{!}{%
\begin{tabular*}{\textwidth}{@{\extracolsep{\fill}}lcccccccccccc@{}}
\toprule
\multirow{2}{*}{\textbf{Model}} & \multicolumn{4}{c}{\textbf{Verified}} & \multicolumn{4}{c}{\textbf{Multi.}} & \multicolumn{4}{c}{\textbf{Pro}} \\
\cmidrule(lr){2-5} \cmidrule(lr){6-9} \cmidrule(lr){10-13}
& \textbf{Static} & \textbf{Dyn.} & \textbf{Score} & \textbf{Solve} & \textbf{Static} & \textbf{Dyn.} & \textbf{Score} & \textbf{Solve} & \textbf{Static} & \textbf{Dyn.} & \textbf{Score} & \textbf{Solve} \\
\midrule
Claude-Sonnet-4.6 & \textbf{96.00} & \textbf{89.00} & \textbf{92.50} & \textbf{85.3} & 95.00 & \underline{98.50} & \underline{96.75} & \underline{95.1} & \textbf{83.50} & \textbf{76.50} & \textbf{80.00} & \textbf{66.0} \\
GLM-5.1 & 93.00 & 86.00 & 89.50 & 83.6 & 95.00 & \textbf{98.50} & \textbf{96.75} & \textbf{95.1} & 79.50 & 59.50 & 69.50 & 55.2 \\
GPT-5.4 & \underline{94.50} & 87.00 & 90.75 & 84.4 & \textbf{96.50} & 87.50 & 92.00 & 88.5 & 77.50 & 56.50 & 67.00 & \underline{56.2} \\
Kimi-K2.5 & 77.56 & 68.67 & 73.11 & 59.1 & \underline{96.49} & 95.61 & 96.05 & 91.2 & \underline{80.67} & \underline{61.08} & \underline{70.88} & 55.2 \\
MiniMax-M2.7 & 77.00 & 68.50 & 72.75 & 63.1 & 72.00 & 90.00 & 81.00 & 72.1 & 28.00 & 43.50 & 35.75 & 18.2 \\
Qwen-3.5 & 94.00 & \underline{87.50} & \underline{90.75} & \underline{84.4} & 87.50 & 94.50 & 91.00 & 83.6 & 68.50 & 55.50 & 62.00 & 43.3 \\
\bottomrule
\end{tabular*}}
\end{table}

Performance degrades consistently from Verified to Pro across all models. The Multilingual subset has an uneven difficulty profile: Env scores are often higher than Verified (smaller repositories, simpler dependency chains), while CodeImpl scores drop sharply (unfamiliar language semantics and toolchains). TestGen on Multilingual tracks Verified for top-tier models but diverges for weaker ones, due to the difficulty of generating discriminative tests in non-Python ecosystems.

\vspace{1em}
\noindent\textbf{Confidence Intervals.}\label{app:ci}
We report 95\% confidence intervals for all core metrics. For binary metrics (Solve), we use Wilson score intervals. For continuous scores (Score), we use bootstrap percentile intervals with 10{,}000 resamples.

\begin{table}[htbp]
\centering
\caption{\textbf{Per-dataset CodeImpl performance with 95\% CI.} Score uses bootstrap CI; Solve uses Wilson CI.}
\label{tab:ci-development}
\small
\setlength{\tabcolsep}{2.5pt}
\renewcommand{\arraystretch}{1.0}
\resizebox{\textwidth}{!}{%
\begin{tabular*}{\textwidth}{@{\extracolsep{\fill}}lcccccc@{}}
\toprule
\multirow{2}{*}{\textbf{Model}} & \multicolumn{2}{c}{\textbf{Verified}} & \multicolumn{2}{c}{\textbf{Multi.}} & \multicolumn{2}{c}{\textbf{Pro}} \\
\cmidrule(lr){2-3} \cmidrule(lr){4-5} \cmidrule(lr){6-7}
& \textbf{Score} & \textbf{Solve} & \textbf{Score} & \textbf{Solve} & \textbf{Score} & \textbf{Solve} \\
\midrule
Claude-Sonnet-4.6 & 72.00{\scriptsize$\pm$4.50} & 53.3{\scriptsize$\pm$6.5} & 58.25{\scriptsize$\pm$9.75} & 34.4{\scriptsize$\pm$11.6} & 53.25{\scriptsize$\pm$4.75} & 28.1{\scriptsize$\pm$6.1} \\
GLM-5.1 & 69.00{\scriptsize$\pm$4.50} & 48.0{\scriptsize$\pm$6.5} & 56.25{\scriptsize$\pm$9.00} & 29.5{\scriptsize$\pm$11.2} & 50.75{\scriptsize$\pm$5.00} & 30.5{\scriptsize$\pm$6.3} \\
GPT-5.4 & 70.00{\scriptsize$\pm$4.50} & 52.0{\scriptsize$\pm$6.5} & 54.00{\scriptsize$\pm$9.75} & 31.1{\scriptsize$\pm$11.3} & 51.75{\scriptsize$\pm$4.75} & 29.1{\scriptsize$\pm$6.2} \\
Kimi-K2.5 & 67.99{\scriptsize$\pm$4.64} & 48.0{\scriptsize$\pm$6.6} & 52.59{\scriptsize$\pm$9.27} & 24.1{\scriptsize$\pm$11.2} & 47.55{\scriptsize$\pm$4.70} & 22.2{\scriptsize$\pm$5.9} \\
MiniMax-M2.7 & 65.00{\scriptsize$\pm$5.00} & 47.1{\scriptsize$\pm$6.5} & 40.25{\scriptsize$\pm$10.25} & 24.6{\scriptsize$\pm$10.6} & 32.00{\scriptsize$\pm$5.00} & 16.3{\scriptsize$\pm$5.1} \\
Qwen-3.5 & 52.25{\scriptsize$\pm$5.50} & 36.4{\scriptsize$\pm$6.2} & 56.50{\scriptsize$\pm$9.50} & 32.8{\scriptsize$\pm$11.5} & 45.00{\scriptsize$\pm$4.50} & 19.2{\scriptsize$\pm$5.4} \\
\bottomrule
\end{tabular*}}
\end{table}

\begin{table}[htbp]
\centering
\caption{\textbf{Per-dataset TestGen performance with 95\% CI.} Score uses bootstrap CI; Solve uses Wilson CI.}
\label{tab:ci-testcase}
\small
\setlength{\tabcolsep}{2.5pt}
\renewcommand{\arraystretch}{1.0}
\resizebox{\textwidth}{!}{%
\begin{tabular*}{\textwidth}{@{\extracolsep{\fill}}lcccccc@{}}
\toprule
\multirow{2}{*}{\textbf{Model}} & \multicolumn{2}{c}{\textbf{Verified}} & \multicolumn{2}{c}{\textbf{Multi.}} & \multicolumn{2}{c}{\textbf{Pro}} \\
\cmidrule(lr){2-3} \cmidrule(lr){4-5} \cmidrule(lr){6-7}
& \textbf{Score} & \textbf{Solve} & \textbf{Score} & \textbf{Solve} & \textbf{Score} & \textbf{Solve} \\
\midrule
Claude-Sonnet-4.6 & 94.75{\scriptsize$\pm$2.00} & 85.8{\scriptsize$\pm$4.6} & 85.25{\scriptsize$\pm$7.00} & 70.5{\scriptsize$\pm$11.2} & 82.50{\scriptsize$\pm$2.75} & 45.8{\scriptsize$\pm$6.8} \\
GLM-5.1 & 90.75{\scriptsize$\pm$2.75} & 77.3{\scriptsize$\pm$5.4} & 87.00{\scriptsize$\pm$6.25} & 68.9{\scriptsize$\pm$11.3} & 76.50{\scriptsize$\pm$3.50} & 39.4{\scriptsize$\pm$6.7} \\
GPT-5.4 & 73.75{\scriptsize$\pm$3.75} & 40.9{\scriptsize$\pm$6.4} & 77.00{\scriptsize$\pm$7.50} & 50.8{\scriptsize$\pm$12.2} & 76.75{\scriptsize$\pm$3.25} & 40.9{\scriptsize$\pm$6.7} \\
Kimi-K2.5 & 87.44{\scriptsize$\pm$2.61} & 63.6{\scriptsize$\pm$6.2} & 78.07{\scriptsize$\pm$7.68} & 54.4{\scriptsize$\pm$12.3} & 76.11{\scriptsize$\pm$3.22} & 38.6{\scriptsize$\pm$6.7} \\
MiniMax-M2.7 & 75.25{\scriptsize$\pm$4.25} & 51.6{\scriptsize$\pm$6.5} & 31.50{\scriptsize$\pm$10.75} & 21.3{\scriptsize$\pm$10.1} & 36.75{\scriptsize$\pm$5.50} & 16.7{\scriptsize$\pm$5.1} \\
Qwen-3.5 & 80.00{\scriptsize$\pm$4.00} & 56.9{\scriptsize$\pm$6.4} & 75.00{\scriptsize$\pm$7.75} & 42.6{\scriptsize$\pm$12.0} & 69.25{\scriptsize$\pm$4.25} & 35.0{\scriptsize$\pm$6.5} \\
\bottomrule
\end{tabular*}}
\end{table}

\begin{table}[htbp]
\centering
\caption{\textbf{Per-dataset Env performance with 95\% CI.} Score uses bootstrap CI; Solve uses Wilson CI.}
\label{tab:ci-environment}
\small
\setlength{\tabcolsep}{2.5pt}
\renewcommand{\arraystretch}{1.0}
\resizebox{\textwidth}{!}{%
\begin{tabular*}{\textwidth}{@{\extracolsep{\fill}}lcccccc@{}}
\toprule
\multirow{2}{*}{\textbf{Model}} & \multicolumn{2}{c}{\textbf{Verified}} & \multicolumn{2}{c}{\textbf{Multi.}} & \multicolumn{2}{c}{\textbf{Pro}} \\
\cmidrule(lr){2-3} \cmidrule(lr){4-5} \cmidrule(lr){6-7}
& \textbf{Score} & \textbf{Solve} & \textbf{Score} & \textbf{Solve} & \textbf{Score} & \textbf{Solve} \\
\midrule
Claude-Sonnet-4.6 & 92.25{\scriptsize$\pm$2.75} & 85.3{\scriptsize$\pm$4.6} & 96.75{\scriptsize$\pm$3.75} & 95.1{\scriptsize$\pm$5.9} & 80.00{\scriptsize$\pm$4.25} & 66.0{\scriptsize$\pm$6.5} \\
GLM-5.1 & 89.50{\scriptsize$\pm$3.50} & 83.6{\scriptsize$\pm$4.8} & 96.75{\scriptsize$\pm$3.75} & 95.1{\scriptsize$\pm$5.9} & 69.50{\scriptsize$\pm$5.00} & 55.2{\scriptsize$\pm$6.8} \\
GPT-5.4 & 91.00{\scriptsize$\pm$3.00} & 84.4{\scriptsize$\pm$4.7} & 92.25{\scriptsize$\pm$5.25} & 88.5{\scriptsize$\pm$8.1} & 67.00{\scriptsize$\pm$5.00} & 56.2{\scriptsize$\pm$6.8} \\
Kimi-K2.5 & 73.11{\scriptsize$\pm$4.89} & 59.1{\scriptsize$\pm$6.2} & 96.05{\scriptsize$\pm$3.95} & 91.2{\scriptsize$\pm$7.0} & 70.88{\scriptsize$\pm$4.90} & 55.2{\scriptsize$\pm$7.2} \\
MiniMax-M2.7 & 73.00{\scriptsize$\pm$5.25} & 63.1{\scriptsize$\pm$6.3} & 81.25{\scriptsize$\pm$8.25} & 72.1{\scriptsize$\pm$11.0} & 35.75{\scriptsize$\pm$5.00} & 18.2{\scriptsize$\pm$5.3} \\
Qwen-3.5 & 90.75{\scriptsize$\pm$3.00} & 84.4{\scriptsize$\pm$4.7} & 91.00{\scriptsize$\pm$5.00} & 83.6{\scriptsize$\pm$9.2} & 62.00{\scriptsize$\pm$5.25} & 43.3{\scriptsize$\pm$6.8} \\
\bottomrule
\end{tabular*}}
\end{table}

\begin{table}[htbp]
\centering
\caption{\textbf{Per-dataset FullCycle performance with 95\% CI.} Score uses bootstrap CI; Solve uses Wilson CI.}
\label{tab:ci-fullcycle}
\small
\setlength{\tabcolsep}{2.5pt}
\renewcommand{\arraystretch}{1.0}
\resizebox{\textwidth}{!}{%
\begin{tabular*}{\textwidth}{@{\extracolsep{\fill}}lcccccc@{}}
\toprule
\multirow{2}{*}{\textbf{Model}} & \multicolumn{2}{c}{\textbf{Verified}} & \multicolumn{2}{c}{\textbf{Multi.}} & \multicolumn{2}{c}{\textbf{Pro}} \\
\cmidrule(lr){2-3} \cmidrule(lr){4-5} \cmidrule(lr){6-7}
& \textbf{Score} & \textbf{Solve} & \textbf{Score} & \textbf{Solve} & \textbf{Score} & \textbf{Solve} \\
\midrule
Claude-Sonnet-4.6 & 83.50{\scriptsize$\pm$2.00} & 20.4{\scriptsize$\pm$5.2} & 82.50{\scriptsize$\pm$3.17} & 8.2{\scriptsize$\pm$7.1} & 76.58{\scriptsize$\pm$1.92} & 4.4{\scriptsize$\pm$2.9} \\
GLM-5.1 & 85.50{\scriptsize$\pm$1.67} & 22.2{\scriptsize$\pm$5.4} & 82.75{\scriptsize$\pm$3.42} & 8.2{\scriptsize$\pm$7.1} & 76.67{\scriptsize$\pm$3.08} & 5.4{\scriptsize$\pm$3.2} \\
GPT-5.4 & 83.92{\scriptsize$\pm$1.67} & 15.1{\scriptsize$\pm$4.7} & 79.00{\scriptsize$\pm$4.75} & 9.8{\scriptsize$\pm$7.6} & 75.83{\scriptsize$\pm$2.42} & 6.4{\scriptsize$\pm$3.4} \\
Kimi-K2.5 & 68.42{\scriptsize$\pm$2.47} & 0.4{\scriptsize$\pm$0.7} & 76.13{\scriptsize$\pm$5.30} & 10.2{\scriptsize$\pm$7.6} & 71.24{\scriptsize$\pm$2.40} & 1.6{\scriptsize$\pm$1.9} \\
MiniMax-M2.7 & 49.00{\scriptsize$\pm$5.17} & 8.4{\scriptsize$\pm$3.7} & 43.00{\scriptsize$\pm$8.75} & 1.6{\scriptsize$\pm$4.2} & 28.58{\scriptsize$\pm$4.83} & 0.5{\scriptsize$\pm$1.3} \\
Qwen-3.5 & 77.83{\scriptsize$\pm$2.58} & 12.4{\scriptsize$\pm$4.3} & 74.42{\scriptsize$\pm$5.50} & 3.3{\scriptsize$\pm$5.1} & 70.67{\scriptsize$\pm$2.42} & 1.5{\scriptsize$\pm$1.9} \\
\bottomrule
\end{tabular*}}
\end{table}

%% file: Appendix/rq2_efficiency.tex
\section{Efficiency Analysis}\label{app:rq2_efficiency}

Table~\ref{tab:rq2-efficiency} reports median output tokens, agent solve time, and evaluation time across all task types. FullCycle requires substantially more tokens and time than isolated tasks. Claude-Sonnet-4.6 achieves the best performance (Table~\ref{tab:isolated-main}) with moderate token consumption and fast execution. Kimi-K2.5 shows moderate solve times (10--30 minutes) with higher token consumption than most models, reflecting a thorough exploration strategy. Evaluation time is stable across models (2--9 minutes for isolated tasks, 6--16 minutes for FullCycle).

\begin{table}[htbp]
\centering
\caption{\textbf{Efficiency metrics across task types.} OutTok = median output tokens (K); Solve = median agent execution time (min); Eval = median verifier execution time (min). All values are medians computed over all instances across three datasets.}
\label{tab:rq2-efficiency}
\small
\setlength{\tabcolsep}{2.5pt}
\renewcommand{\arraystretch}{1.0}
\resizebox{\textwidth}{!}{%
\begin{tabular*}{\textwidth}{@{\extracolsep{\fill}}lcccccccccccc@{}}
\toprule
\multirow{2}{*}{\textbf{Model}} & \multicolumn{3}{c}{\textbf{Impl}} & \multicolumn{3}{c}{\textbf{TestGen}} & \multicolumn{3}{c}{\textbf{Env}} & \multicolumn{3}{c}{\textbf{FullCycle}} \\
\cmidrule(lr){2-4} \cmidrule(lr){5-7} \cmidrule(lr){8-10} \cmidrule(lr){11-13}
& \textbf{OutTok} & \textbf{Solve} & \textbf{Eval} & \textbf{OutTok} & \textbf{Solve} & \textbf{Eval} & \textbf{OutTok} & \textbf{Solve} & \textbf{Eval} & \textbf{OutTok} & \textbf{Solve} & \textbf{Eval} \\
\midrule
Claude-Sonnet-4.6 & 5.4K & 5.7 & 3.6 & 7.3K & 8.4 & 3.9 & 3.1K & 8.2 & 2.9 & 11.0K & 11.7 & 8.0 \\
GLM-5.1 & 3.0K & 11.6 & 3.1 & 3.0K & 12.1 & 4.0 & 1.5K & 12.7 & 2.6 & 5.9K & 13.8 & 7.7 \\
GPT-5.4 & 2.8K & 4.2 & 3.2 & 2.6K & 4.9 & 4.0 & 4.4K & 20.1 & 2.6 & 6.0K & 10.9 & 8.6 \\
Qwen-3.5 & 8.9K & 23.5 & 3.1 & 6.7K & 18.8 & 3.9 & 2.6K & 21.9 & 2.7 & 12.4K & 29.5 & 7.6 \\
Kimi-K2.5 & 8.5K & 10.8 & 4.0 & 7.5K & 10.4 & 5.0 & 4.0K & 29.5 & 3.5 & 10.6K & 27.8 & 16.0 \\
MiniMax-M2.7 & 4.2K & 12.3 & 3.2 & 5.9K & 16.1 & 3.8 & 2.0K & 14.1 & 2.8 & 1.9K & 5.7 & 6.0 \\
\bottomrule
\end{tabular*}}
\end{table}

%% file: Appendix/rq3_script_failures.tex
\section{Script Evaluation Failure Analysis}\label{app:disagree_cases}

\subsection{Disagreement Annotation Protocol}

To categorize Script--SWE-Judge disagreements, two graduate researchers with software engineering experience independently label each case. For each disagreement instance, annotators receive a review package containing: the issue description, the agent's submitted patch, the gold reference patch, SWE-Judge's scoring with reasoning, the script evaluator's binary verdict with execution logs, and LLM-generated auxiliary analysis highlighting potential discrepancies.

Each annotation follows a structured protocol:
\begin{enumerate}[leftmargin=15pt]
    \item Read the issue description to understand the problem context.
    \item Examine the gold reference patch to establish the correct solution approach.
    \item Review the agent's submission to understand what the agent implemented.
    \item Read SWE-Judge's scoring and reasoning.
    \item Cross-reference with execution logs and LLM auxiliary analysis when static review is insufficient.
    \item Assign a failure category from the predefined taxonomy and record which evaluator is correct.
\end{enumerate}

\paragraph{Human Verification Results.}
To validate the LLM-assisted categorization and rule out selection bias, we conduct human deep annotation on all 371 disagreement instances plus 86 agreement instances (a 10\% random sample of cases where SWE-Judge and the script concur). For disagreement cases, human annotators confirm SWE-Judge as correct in 98.6\% (366/371), the script as correct in 0.5\% (2/371), and neither in 0.8\% (3/371). For agreement cases, 97.8\% (85/86) are confirmed correct by human review, indicating that evaluator consensus reliably reflects ground truth with negligible risk of systematic co-failure.

\subsection{Disagreement Categories}

Table~\ref{tab:rq3-errors} summarizes the categorization results.

\begin{table}[htbp]
\centering
\caption{Script Evaluation Errors from 3,267 Script--SWE-Judge Disagreements. \% denotes the proportion of each category.}
\label{tab:rq3-errors}
\begin{tabular}{lcc}
\toprule
\textbf{Category} & \textbf{N} & \textbf{\%} \\
\midrule
Excessive strictness     & 1176 & 36.0 \\
Evaluation breakdown     & 1072 & 32.8 \\
Excessive leniency       &  882 & 27.0 \\
Others                   &  137 &  4.2 \\
\bottomrule
\end{tabular}
\end{table}

\subsection{Representative Case Studies}

We present representative cases organized by the three failure categories.

\subsection{Excessive Strictness}\label{app:strict}

Scripts demand exact alignment with the golden patch and reject functionally equivalent alternatives. Two manifestations appear: alternative implementations receiving zero credit, and partial fixes losing all information through binary scoring.

\paragraph{Case 1: Alternative Implementation in Valkey.}
In \texttt{valkey-io/valkey\#1499}, the golden patch modifies the command table to fix a permission checking issue. The agent instead uses \texttt{executing\_client->cmd} to check the actual command being executed, with null-safety handling. All \textsc{fail\_to\_pass} tests pass and no \textsc{pass\_to\_pass} regressions occur. The script assigns 0 because its tests are coupled to the specific implementation path of the golden patch. SWE-Judge performs static analysis, confirms the semantic equivalence of both approaches, and assigns 1.0.

\paragraph{Case 2: Constant Exporting in Teleport.}
In a Teleport issue requiring namespace configuration constants, the golden patch inlines string literals across multiple files. The agent exports \texttt{NamespaceEnv} and \texttt{ReleaseNameEnv} as package-level constants and updates all references. This is a cleaner refactoring that produces identical behavior. All 13 \textsc{fail\_to\_pass} tests pass, but the script assigns 0 because the modified file set differs from the expected set. SWE-Judge recognizes the functional equivalence and awards full marks.

\paragraph{Case 3: Partial Fix in PHPSpreadsheet.}
In a PHPSpreadsheet task, the agent correctly adds a \texttt{\_\_toString()} method to the \texttt{StructuredReference} class, fixing the immediate string conversion error. However, it misses additional changes for cross-worksheet table and structured reference handling that the golden patch includes. The script assigns 0, indistinguishable from a completely wrong submission. SWE-Judge assigns 0.25, recognizing that the core direction is correct but coverage is incomplete. This proportional credit separates near-miss attempts from zero-effort submissions.

\paragraph{Case 4: Qutebrowser Path Resolution.}
The agent creates \texttt{FilePathCategory} with proper path resolution for \texttt{file://}, tilde, and absolute paths, integrates it into the URL model, and updates documentation. The implementation handles all major scenarios but misses minor edge cases in the golden patch. SWE-Judge assigns 0.75, reflecting a nearly complete solution. The script's binary 0 fails to capture this meaningful progress.

\subsection{Evaluation Breakdown}\label{app:breakdown}

Nearly a third of disagreements occur because the evaluation pipeline itself fails, independent of solution quality. Parser incompatibilities and infrastructure rot are the primary causes.

\paragraph{Case 5: Gradle Output Parsing in Apache Lucene.}
Multiple Apache Lucene environment tasks exhibit this pattern. The agent correctly configures JDK~21, the Gradle wrapper, and all build dependencies. SWE-Judge independently confirms via JUnit XML that all 108 tests pass with 0 failures. However, the SWE-bench evaluation parser expects Maven-style output format and cannot parse Gradle's \texttt{BUILD SUCCESSFUL} format, reporting a zero score. This is a failure of the evaluation tool, not the agent.

\paragraph{Case 6: Maven Daemon Timeout in Google Gson.}
In a Google Gson environment task, the agent's setup correctly installs Java and Maven. All 10 tests pass when executed with standard Maven. However, the evaluation script uses \texttt{mvnd} (Maven Daemon), which times out during cold start. SWE-Judge identifies this as an infrastructure artifact: the original script evaluation failures stem from \texttt{mvnd} daemon issues (timeout/crashes), not actual test failures. The agent receives full marks from SWE-Judge.

\paragraph{Case 7: Node.js Workspace Corruption.}
After \texttt{yarn install}, the \texttt{node\_modules} state file is missing or corrupted in approximately 178 cases, causing all subsequent commands to fail with ``Couldn't find the node\_modules state file.'' The agent's code is never evaluated because the test framework collapses before reaching any relevant assertion. SWE-Judge evaluates the agent's configuration through static analysis and awards credit based on the quality of the submitted patch, independent of whether the test infrastructure executed successfully.

These cases illustrate a structural limitation: script evaluation conflates ``the framework crashed'' with ``the solution is wrong.'' As dependencies deprecate and runtime versions drift, this conflation worsens over time.

\subsection{Excessive Leniency}\label{app:lenient}

Script evaluation can conflate superficial execution success with semantic correctness.

\paragraph{Case 8: Trivial State Transition in TestGen.}
In an Ansible test task, the agent writes tests that import \texttt{set\_multipart\_encoding} at module level. This function exists only after the fix. On buggy code, the test fails with \texttt{ImportError} before any test logic executes. The script's dynamic state transition protocol checks only whether Phase~1 (buggy code) produces a non-zero exit code and Phase~2 (fixed code) passes. Both conditions are met, so the script awards full marks. SWE-Judge recognizes the failure mechanism: ``When \texttt{code\_patch} is reverted, module import fails with \texttt{AttributeError} before any tests can run to detect actual bug behavior.'' The test provides zero discriminative power because any pre-fix version would fail regardless of the specific bug.

\paragraph{Case 9: Regression Escape in Django.}
In \texttt{django/django\#13590}, the agent correctly fixes \texttt{namedtuple} support in \texttt{Values()} by unpacking values. However, the patch unconditionally unpacks all types, breaking regular list/tuple construction. The \textsc{fail\_to\_pass} test passes, but 5 \textsc{pass\_to\_pass} tests fail with \texttt{TypeError}. The script monitors only the target test scope and awards full marks. SWE-Judge runs the complete test suite and identifies the regression: the golden patch uses \texttt{hasattr(type\_, '\_make')} to detect namedtuples and only unpacks for those types. SWE-Judge assigns 0.25.

\paragraph{Case 10: Incomplete Test Coverage in coreutils.}
In \texttt{uutils/coreutils\#6575} (TestGen), the agent tests non-UTF-8 filename handling but covers only CRC mode, missing the SHA256 mode test present in the golden patch. The script's pass/fail check accepts this single-scenario test as fully correct. SWE-Judge evaluates coverage depth against the golden patch and assigns 0.25, recognizing that the test provides insufficient coverage to serve as a reliable regression test.

\paragraph{Case 11: Testing Unchanged Code in Vuls.}
In a Vuls test task, the agent modifies an existing test to call \texttt{convertToModel()} (which is unchanged between buggy and fixed states) instead of testing the actual bug in \texttt{config/os.go}. The test passes in both states, providing zero discriminative power. The script awards full marks based on the exit code. SWE-Judge compares the test logic against the issue description and golden patch, identifying that the tested function is irrelevant to the reported bug. Score: 0.25.

%% file: Appendix/rq3_judge_workflow.tex
\section{SWE-Judge Workflow Case Studies}\label{app:judge_workflow}

We select three FullCycle evaluation workflows from Table~\ref{tab:rq3-behavior} that each illustrate a distinct capability: adaptive eval scripting, fault injection, and build verification with multi-dimensional scoring.

\subsection{Case 1: Adaptive Eval Scripting}\label{app:workflow_adaptive}

\paragraph{Instance.} \texttt{NodeBB/NodeBB\#8168c6c4} (FullCycle, Claude-Sonnet-4.6). The issue requires implementing profile image cleanup: when users remove cover photos or avatars, the corresponding files on disk must be deleted.

\paragraph{Workflow Summary.} SWE-Judge executes 6 steps across 47 tool calls:

\begin{enumerate}[leftmargin=10pt]
\item \textbf{Instruction and gold patch review.} SWE-Judge reads the issue description and golden patch to establish the expected behavior: file deletion via \texttt{rimraf} with glob patterns for accumulated profile images.

\item \textbf{Agent patch macro-review.} SWE-Judge reviews the agent's diff across 5 modified files (\texttt{src/groups/cover.js}, \texttt{src/socket.io/user/picture.js}, \texttt{src/user/delete.js}, \texttt{src/user/picture.js}). It identifies a critical divergence: the agent uses \texttt{getLocalCoverPath}/\texttt{getLocalAvatarPath} to delete only the current file, while the golden patch uses glob patterns to delete all accumulated files.

\item \textbf{ENV evaluation.} Static: \texttt{setup.sh} runs \texttt{npm install} correctly. Dynamic: Node.js v18.20.8 available, packages import successfully, 359 tests collected. Score: 4/4.

\item \textbf{TEST evaluation.} Static: Agent covers 3 of 4 key scenarios (missing account deletion cleanup test). Dynamic: Phase~1 fails with \texttt{TypeError: User.getLocalCoverPath is not a function} (imprecise failure). Score: 2/4.

\item \textbf{Adaptive eval scripting (triggered by TEST\_STATIC < 2).} The agent's test suite lacks coverage for account deletion cleanup. SWE-Judge writes \texttt{eval\_improved.sh} and \texttt{test/improved-image-cleanup.js}, a custom test that creates dummy profile images, calls the account deletion function, and verifies that 0 files remain afterward. The first execution discovers 3 orphaned files. After debugging a path configuration issue and re-executing, the test confirms the agent's implementation leaves orphaned files during account deletion.

\item \textbf{CODE evaluation using custom test results.} The custom test output directly informs CODE\_DYNAMIC: 3/4 tests pass (group cover, user cover, user avatar succeed; account deletion cleanup fails). Score: 2/4.
\end{enumerate}

\paragraph{Final Scores.} ENV: 4, TEST: 2, CODE: 2. Total: 8/12 (0.667).

This case shows that SWE-Judge writes its own verification scripts when existing coverage is insufficient (34.6\% of FullCycle evaluations), exposing gaps that the agent's own tests miss.

\subsection{Case 2: Fault Injection}\label{app:workflow_fault}

\paragraph{Instance.} \texttt{internetarchive/openlibrary} (FullCycle, Qwen-3.5). The issue requires adding a Solr boolean clause limit configuration (\texttt{-Dsolr.max.booleanClauses=30000}) to \texttt{docker-compose.yml} and a corresponding \texttt{FILTER\_BOOK\_LIMIT} constant in \texttt{bookshelves.py}.

\paragraph{Workflow Summary.}

\begin{enumerate}[leftmargin=10pt]
\item \textbf{Initial review.} SWE-Judge reads the instruction, golden patch, and agent patch. The agent correctly implements both required changes: adding \texttt{-Dsolr.max.booleanClauses=30000} to \texttt{SOLR\_OPTS} and defining \texttt{FILTER\_BOOK\_LIMIT = 30\_000}.

\item \textbf{Agent test execution (Phase 2).} SWE-Judge runs the agent's test suite (\texttt{eval.sh}) on the fixed code. Both tests pass: \texttt{test\_filter\_book\_limit\_constant\_exists} and \texttt{test\_solr\_opts\_has\_boolean\_clauses\_limit}.

\item \textbf{Fault injection (Phase 1).} SWE-Judge reverts the agent's changes to simulate the buggy state:
\begin{verbatim}
git show base_commit:docker-compose.yml > /tmp/docker-compose-buggy.yml
cp /tmp/docker-compose-buggy.yml docker-compose.yml
git show base_commit:openlibrary/core/bookshelves.py > \
    /tmp/bookshelves-buggy.py
cp /tmp/bookshelves-buggy.py openlibrary/core/bookshelves.py
\end{verbatim}
SWE-Judge then re-runs the agent's tests against this reverted codebase. Both tests now fail: \texttt{FILTER\_BOOK\_LIMIT} is not found in \texttt{bookshelves.py}, and \texttt{-Dsolr.max.booleanClauses} is absent from \texttt{SOLR\_OPTS}.

\item \textbf{Verdict.} The tests correctly discriminate between buggy and fixed states. SWE-Judge confirms the agent's tests are not trivial or overfitted: they verify specific code content rather than relying on indirect signals. TEST\_DYNAMIC: 2/2.

\item \textbf{ENV evaluation.} Static: Agent uses venv instead of the requested conda environment, deviating from the instruction. Dynamic: Python 3.11.1 available, but core package import fails (\texttt{ModuleNotFoundError: web}). Score: 2/4.
\end{enumerate}

\paragraph{Final Scores.} ENV: 2, TEST: 3, CODE: 4. Total: 9/12 (0.75).

Fault injection verifies that the agent's tests genuinely detect the bug rather than passing for spurious reasons. SWE-Judge uses this technique in 4.8\% of evaluations, typically when the tests appear suspiciously simple or when configuration changes could easily produce false positives.

\subsection{Case 3: Build Verification and Multi-Dimensional Scoring}\label{app:workflow_build}

\paragraph{Instance.} \texttt{flipt-io/flipt\#292fdac} (FullCycle, Claude-Sonnet-4.6). The issue requires implementing an optional configuration versioning feature for the Flipt feature flag server (Go).

\paragraph{Workflow Summary.}

\begin{enumerate}[leftmargin=10pt]
\item \textbf{Code review via git diff.} SWE-Judge examines the agent's changes: adding a \texttt{Version} field to the configuration struct, implementing validation logic, updating the schema, and creating test data files.

\item \textbf{Reference comparison.} SWE-Judge reads the golden patch and performs a structural comparison. The agent's implementation aligns closely with the golden patch, using cleaner error handling patterns in some cases.

\item \textbf{Build verification.} SWE-Judge runs \texttt{go build ./...} to confirm compilation succeeds, then uses a non-matching test pattern to verify test collection without execution.

\item \textbf{Test execution with fault injection.} SWE-Judge reverts the code to the buggy state and runs the agent's tests. Tests fail with \texttt{cfg.Version undefined} (compilation error). SWE-Judge notes this is a weaker detection mechanism (compile-time rather than assertion-based) but still validates that the tests cannot pass without the fix.

\item \textbf{Multi-dimensional scoring.}
\begin{itemize}
\item ENV: Static 2/2 (complete setup), Dynamic 2/2 (Go toolchain available, packages import, tests collect). Score: 4/4.
\item TEST: Static 2/2 (comprehensive test coverage aligned with golden patch), Dynamic 1/2 (Phase~1 failure is imprecise: compilation error rather than assertion failure). Score: 3/4.
\item CODE: Static 2/2 (correct implementation matching golden patch), Dynamic 2/2 (all target tests pass on fixed code). Score: 4/4.
\end{itemize}
\end{enumerate}

\paragraph{Final Scores.} ENV: 4, TEST: 3, CODE: 4. Total: 11/12 (0.917).

Build verification (used in 36.1\% of FullCycle evaluations, mostly compiled languages) serves as a gate: a failed build immediately invalidates dynamic scores. The multi-dimensional scoring here separates a correct implementation (CODE: 4/4) from an imprecise test design (TEST: 3/4), a distinction that binary pass/fail cannot express.

%% file: Appendix/ablation_blind.tex
\section{Ablation: Reference-Guided vs.\ Blind Evaluation}
\label{app:appendix_blind}

Does access to the official patch cause the evaluator to penalize valid alternative implementations? We compare \textbf{Gold eval} (evaluator receives the reference solution) against \textbf{Blind eval} (evaluator judges solely from the problem description, repository state, and submitted patch).

\subsection{Score Comparison}
\label{app:blind-scores}

Table~\ref{tab:app-ablation-combined} reports the dimension-level comparison across 8{,}678 paired trials.

\begin{table}[htbp]
\centering
\caption{Gold vs.\ Blind evaluation comparison across all task categories. Isolated: $n{=}5{,}771$; FullCycle: $n{=}2{,}907$. All scores normalized to percentages. Diff = Blind $-$ Gold (percentage points).}
\label{tab:app-ablation-combined}
\begin{tabular}{llccc}
\toprule
\textbf{Setting} & \textbf{Category} & \textbf{Gold (\%)} & \textbf{Blind (\%)} & \textbf{Diff (pp)} \\
\midrule
\multirow{2}{*}{Isolated} & Impl    & 64.0 & 68.4 & +4.4 \\
                              & TestGen & 64.1 & 66.4 & +2.3 \\
\midrule
\multirow{3}{*}{FullCycle}    & ENV     & 83.2 & 82.5 & $-$0.7 \\
                              & CODE    & 64.3 & 82.7 & \textbf{+18.4} \\
                              & TEST    & 62.9 & 68.1 & +5.3 \\
\bottomrule
\end{tabular}
\end{table}

In Isolated evaluation, Blind eval inflates Impl scores by +4.4\,pp and TestGen by +2.3\,pp. In FullCycle, the inflation concentrates in CODE (+18.4\,pp), where assessing correctness without a reference is hardest. ENV scores remain stable ($-$0.7\,pp) because environment correctness is largely verifiable through execution. The inflation is driven almost entirely by static sub-scores ($\Delta_S{=}+0.17$) while dynamic sub-scores remain unchanged ($\Delta_D{=}+0.01$): execution-based verification is objective regardless of reference availability.


\subsection{False Negative Analysis}
\label{app:blind-fn}

If reference access biased against correct alternatives, Gold eval would show an elevated false negative (FN) rate. Table~\ref{tab:app-error-rates} measures this using the script evaluator as ground truth (threshold 0.5).

\begin{table}[htbp]
\centering
\caption{Error rates by category (binary threshold = 0.5). FP = false positive (incorrect submission scored as pass). FN = false negative (correct submission scored as fail).}
\label{tab:app-error-rates}
\begin{tabular}{lcccc}
\toprule
\textbf{Category} & \textbf{Gold FP} & \textbf{Blind FP} & \textbf{Gold FN} & \textbf{Blind FN} \\
\midrule
Impl    & 8.7\%  & 14.3\% & 0.3\% & 0.2\% \\
TestGen & 35.2\% & 34.9\% & 0.2\% & 0.2\% \\
\bottomrule
\end{tabular}
\end{table}

Gold eval's FN rate is $\leq$0.3\% across both categories, virtually identical to Blind eval. Providing the reference does not cause the evaluator to reject valid submissions. The measurable difference is in false positives: Blind eval's FP rate on Impl is 1.6$\times$ that of Gold (14.3\% vs.\ 8.7\%). The reference improves precision without increasing rigidity.

\subsection{Illustrative Case: Gold Eval Favors a Correct Alternative}
\label{app:blind-case}

We present a case where Gold eval is more lenient than Blind eval toward an alternative implementation.

\medskip
\noindent\textbf{\texttt{django\_\_django-16877}} (Impl, Claude 4.6) --- Script=0, Gold=4/4, Blind=1/4

\begin{quote}
\textit{Gold (static=2):} ``Implementation is functionally identical to gold.patch---correctly implements \texttt{escapeseq} filter with equivalent logic.''

\textit{Blind (static=1):} ``Fix direction correct, but agent left unresolved merge conflicts in test file.''
\end{quote}

\noindent Gold eval confirms semantic equivalence with the reference and correctly identifies the merge conflict markers as irrelevant to functional correctness. Blind eval, lacking this anchor, is misled by the cosmetic issue and penalizes a correct submission. Reference access here protects the alternative implementation by providing a semantic equivalence check.

\subsection{Difficulty Stratification}
\label{app:blind-difficulty}

Table~\ref{tab:app-difficulty} stratifies trials by Gold score to examine where the Gold--Blind gap concentrates.

\begin{table}[htbp]
\centering
\caption{Difficulty stratification: Blind--Gold score difference by Gold score bin ($n{=}5{,}771$ Isolated trials).}
\label{tab:app-difficulty}
\begin{tabular}{lrcr}
\toprule
\textbf{Gold Score Bin} & \textbf{$n$} & \textbf{Blind Mean} & \textbf{Diff} \\
\midrule
$=0$ (clearly wrong)     & 1{,}017 & 0.024 & +0.024 \\
$(0, 0.25]$              & 1{,}033 & 0.289 & +0.039 \\
$(0.25, 0.5]$            & 388     & 0.599 & +0.099 \\
$(0.5, 0.75]$ (partial)  & 1{,}033 & 0.886 & \textbf{+0.136} \\
$(0.75, 1.0]$ (correct)  & 2{,}300 & 0.978 & $-$0.022 \\
\bottomrule
\end{tabular}
\end{table}

The pattern forms an inverted-U: inflation peaks at the $(0.5, 0.75]$ bin (+0.136) and reverses for near-perfect submissions ($-$0.022). If Gold eval penalized correct alternatives, the highest-scoring bin would show Gold $>$ Blind. Instead, the slight negative difference confirms Gold eval does not under-score correct submissions. The only divergence direction is Blind eval over-scoring partial fixes in the ambiguous middle range.

%% file: Appendix/rq4_details.tex
\section{End-to-End vs.\ Isolated Details}\label{app:rq4}

Per-model breakdowns supporting Section~\ref{sec:rq4}. All bonus counts exclude timeout-driven flips.

\subsection{Instance-Level Flip Counts}
\label{app:rq4-flip}

\begin{table}[htbp]
\centering
\caption{Per-model instance-level flips between FullCycle and Isolated ($N{=}489$ instances per model). \textbf{Degrad.} = instances solved in Isolated but imperfect in FullCycle. \textbf{Bonus} = instances unsolved in Isolated but perfect in FullCycle (timeout-driven cases excluded).}
\label{tab:rq4-flip-per-model}
\resizebox{\textwidth}{!}{%
\begin{tabular}{l|cc|cc|cc}
\toprule
 & \multicolumn{2}{c|}{\textbf{Env}} & \multicolumn{2}{c|}{\textbf{CodeImpl}} & \multicolumn{2}{c}{\textbf{TestGen}} \\
\cmidrule(lr){2-3} \cmidrule(lr){4-5} \cmidrule(lr){6-7}
\textbf{Model} & Degrad. & Bonus & Degrad. & Bonus & Degrad. & Bonus \\
\midrule
Claude-Sonnet-4.6     & 73  & 31 & 92  & 23 & 193 & 27 \\
GLM-5.1        & 51  & 75 & 83  & 33 & 158 & 40 \\
GPT-5.4        & 55  & 59 & 101 & 46 & 111 & 56 \\
Kimi-K2.5      & 169 & 31 & 78  & 26 & 177 & 26 \\
MiniMax-M2.7   & 112 & 67 & 95  & 9  & 134 & 17 \\
Qwen-3.5       & 42  & 64 & 81  & 28 & 149 & 26 \\
\midrule
\textbf{Total} & 356 & 301 & 519 & 152 & 815 & 177 \\
\bottomrule
\end{tabular}}
\end{table}

The pipeline gradient holds across all models: degradation counts increase monotonically from Env to TestGen, confirming downstream dimensions bear a heavier integration tax. MiniMax-M2.7 shows the most severe net Env degradation. GLM-5.1 achieves a net Env bonus (75 vs.\ 51), indicating effective use of downstream signals to repair environment defects. For CodeImpl, GPT-5.4 has the highest bonus count (46) but also the highest degradation (101). Even the best models (Claude-Sonnet-4.6: 193 degradation vs.\ 27 bonus) suffer a roughly 7:1 degradation-to-bonus ratio in TestGen.

\subsection{Degradation Root Cause Analysis}
\label{app:rq4-cause}

\begin{table}[htbp]
\centering
\caption{Degradation root cause distribution per model and dimension. \textbf{S-only} = static score drops while dynamic remains perfect. \textbf{Both} = both static and dynamic degrade. Percentages are computed over score-related cases only (excluding timeout).}
\label{tab:rq4-deg-cause}
\resizebox{\textwidth}{!}{%
\begin{tabular}{l|ccc|ccc|ccc}
\toprule
 & \multicolumn{3}{c|}{\textbf{Env}} & \multicolumn{3}{c|}{\textbf{CodeImpl}} & \multicolumn{3}{c}{\textbf{TestGen}} \\
\cmidrule(lr){2-4} \cmidrule(lr){5-7} \cmidrule(lr){8-10}
\textbf{Model} & S-only & Both & Total & S-only & Both & Total & S-only & Both & Total \\
\midrule
Claude-Sonnet-4.6     & 62\,(89\%) & 7\,(10\%)  & 70  & 83\,(92\%) & 7\,(8\%)   & 90  & 138\,(74\%) & 37\,(20\%) & 187 \\
GLM-5.1        & 35\,(70\%) & 14\,(28\%) & 50  & 79\,(98\%) & 1\,(1\%)   & 81  & 116\,(74\%) & 32\,(20\%) & 157 \\
GPT-5.4        & 30\,(58\%) & 19\,(37\%) & 52  & 86\,(85\%) & 14\,(14\%) & 101 & 65\,(59\%)  & 38\,(35\%) & 110 \\
Kimi-K2.5      & 117\,(76\%) & 37\,(24\%) & 154 & 60\,(72\%) & 23\,(28\%) & 83  & 120\,(54\%) & 101\,(46\%) & 221 \\
MiniMax-M2.7   & 22\,(22\%) & 77\,(77\%) & 100 & 32\,(36\%) & 56\,(64\%) & 88  & 43\,(33\%)  & 88\,(67\%) & 132 \\
Qwen-3.5       & 25\,(61\%) & 15\,(37\%) & 41  & 63\,(79\%) & 17\,(21\%) & 80  & 91\,(61\%)  & 55\,(37\%) & 148 \\
\bottomrule
\end{tabular}}
\end{table}

Table~\ref{tab:rq4-deg-cause} shows a capability divide. For CodeImpl, SOTA models (Claude-Sonnet-4.6, GLM-5.1, GPT-5.4) degrade almost exclusively through static-only loss (85--98\%): their code runs correctly but sacrifices structural quality during iterative patching. MiniMax-M2.7 shows the opposite, with 64\% joint collapse in CodeImpl and 77\% in Env, meaning weaker models cannot maintain functional correctness under integration pressure. The same pattern holds for TestGen: SOTA models show 59--74\% static-only loss (tests execute but coverage drops), while MiniMax-M2.7 suffers 67\% joint failure. Kimi-K2.5 follows the SOTA pattern in Env (76\% static-only) and CodeImpl (72\% static-only), but shows a higher joint collapse rate in TestGen (46\%), suggesting that its strong isolated TestGen performance degrades more under integration pressure.

\subsection{Static vs.\ Dynamic Score Comparison}
\label{app:rq4-sd}

\begin{table}[htbp]
\centering
\caption{Static and Dynamic sub-scores (0--1 scale) in Isolated (ISO) vs.\ FullCycle (FC). The CodeImpl dimension exhibits a stark reversal: Static declines while Dynamic surges, driven by the write-run-fix loop.}
\label{tab:rq4-sd-comparison}
\resizebox{\textwidth}{!}{%
\begin{tabular}{l|cccc|cccc|cccc}
\toprule
 & \multicolumn{4}{c|}{\textbf{Env}} & \multicolumn{4}{c|}{\textbf{CodeImpl}} & \multicolumn{4}{c}{\textbf{TestGen}} \\
\cmidrule(lr){2-5} \cmidrule(lr){6-9} \cmidrule(lr){10-13}
\textbf{Model} & S\textsubscript{ISO} & S\textsubscript{FC} & D\textsubscript{ISO} & D\textsubscript{FC} & S\textsubscript{ISO} & S\textsubscript{FC} & D\textsubscript{ISO} & D\textsubscript{FC} & S\textsubscript{ISO} & S\textsubscript{FC} & D\textsubscript{ISO} & D\textsubscript{FC} \\
\midrule
Claude-Sonnet-4.6     & 0.91 & 0.85 & 0.85 & 0.98 & 0.68 & 0.61 & 0.57 & 0.90 & 0.91 & 0.69 & 0.86 & 0.81 \\
GLM-5.1        & 0.88 & 0.88 & 0.76 & 0.97 & 0.66 & 0.62 & 0.54 & 0.89 & 0.86 & 0.70 & 0.83 & 0.83 \\
GPT-5.4        & 0.88 & 0.87 & 0.74 & 0.93 & 0.69 & 0.63 & 0.52 & 0.89 & 0.74 & 0.66 & 0.76 & 0.81 \\
Kimi-K2.5      & 0.81 & 0.66 & 0.69 & 0.90 & 0.64 & 0.59 & 0.51 & 0.81 & 0.81 & 0.56 & 0.82 & 0.71 \\
MiniMax-M2.7   & 0.56 & 0.46 & 0.61 & 0.56 & 0.53 & 0.32 & 0.43 & 0.43 & 0.53 & 0.26 & 0.55 & 0.36 \\
Qwen-3.5       & 0.83 & 0.85 & 0.75 & 0.95 & 0.56 & 0.56 & 0.43 & 0.80 & 0.73 & 0.56 & 0.77 & 0.74 \\
\bottomrule
\end{tabular}}
\end{table}

Table~\ref{tab:rq4-sd-comparison} quantifies the Static-Dynamic reversal from Section~\ref{sec:rq4}:

\textbf{Env.} For SOTA models, Static scores remain stable or decline slightly (Claude: $0.91 \to 0.85$) while Dynamic scores rise (Claude: $0.85 \to 0.98$), because cross-phase runtime feedback catches configuration defects that isolated evaluation misses. Kimi-K2.5 follows the typical pattern with a Static decline ($0.81 \to 0.66$) and a strong Dynamic gain ($0.69 \to 0.90$), consistent with the runtime feedback mechanism. MiniMax-M2.7 is the only model where both Env sub-scores decline.

\textbf{CodeImpl.} The reversal is sharpest here. In Isolated mode, SOTA models show Static $>$ Dynamic (Claude: $0.68 > 0.57$), producing well-structured code that fails at runtime. In FullCycle, this inverts to Dynamic $\gg$ Static (Claude: $0.90 \gg 0.61$). The average Dynamic gain across SOTA models is $+0.35$, while Static declines by only $-0.06$: the write-run-fix loop improves runtime correctness but no equivalent signal guards structural quality. MiniMax-M2.7 shows no Dynamic improvement ($0.43 \to 0.43$) alongside severe Static collapse ($0.53 \to 0.32$).

\textbf{TestGen.} Static scores drop for all models (Claude: $0.91 \to 0.69$, Qwen: $0.73 \to 0.56$), consistent with attention depletion at the pipeline tail. Dynamic scores are mixed: SOTA models maintain or slightly lose ground (Claude: $0.86 \to 0.81$), while GPT-5.4 improves ($0.76 \to 0.81$) through the self-implementation knowledge mechanism. MiniMax-M2.7 collapses on both sub-scores.
